\documentclass[10pt,twocolumn,letterpaper]{article}

\usepackage{iccv}
\usepackage{times}
\usepackage{epsfig}
\usepackage{graphicx}
\usepackage{amsmath}
\usepackage{amssymb}
\usepackage{multirow}
\usepackage{makecell}
\usepackage{float}
\usepackage{appendix}
\usepackage[section]{placeins}
\usepackage[ruled, linesnumbered]{algorithm2e}

\newcommand{\rNum}[1]{\lowercase\expandafter{\romannumeral #1\relax}}
\makeatletter
\newcommand{\removelatexerror}{\let\@latex@error\@gobble}
\makeatother

\usepackage[breaklinks=true,bookmarks=true, bookmarksnumbered=true, bookmarksopen=true]{hyperref}

\iccvfinalcopy 

\def\iccvPaperID{****} 

\ificcvfinal\fi

\begin{document}

\title{Stochastic Image Denoising by Sampling from the Posterior Distribution}

\author{Bahjat Kawar
\qquad
Gregory Vaksman
\qquad
Michael Elad
\\
Computer Science Department, The Technion - Israel Institute of Technology\\
{\tt\small \{bahjat.kawar,grishav,elad\}@cs.technion.ac.il}
}

\maketitle
\ificcvfinal\thispagestyle{empty}\fi

\begin{abstract}
   Image denoising is a well-known and well studied problem, commonly targeting a minimization of the mean squared error (MSE) between the outcome and the original image. Unfortunately, especially for severe noise levels, such Minimum MSE (MMSE) solutions may lead to blurry output images. In this work we propose a novel stochastic denoising approach that produces viable and high perceptual quality results, while maintaining a small MSE. Our method employs Langevin dynamics that relies on a repeated application of any given MMSE denoiser, obtaining the reconstructed image by effectively sampling from the posterior distribution. Due to its stochasticity, the proposed algorithm can produce a variety of high-quality outputs for a given noisy input, all shown to be legitimate denoising results. In addition, we present an extension of our algorithm for handling the inpainting problem, recovering missing pixels while removing noise from partially given data.
\end{abstract}


\section{Introduction}
This work focuses on the image denoising task, a well-known and well-studied problem in the field of image processing. Various successful algorithms, both classically oriented and deep learning based,  were proposed over the years for handling this task, such as NLM~\cite{buades2005non}, KSVD~\cite{elad2006image}, BM3D~\cite{dabov2007image}, EPLL~\cite{zoran2011learning}, WNNM~\cite{gu2014weighted},  TNRD~\cite{chen2016trainable}, DnCNN~\cite{DnCNN}, NLRN~\cite{liu2018non} and others~\cite{yu2011solving,roth2005fields,aharon2008sparse,vaksman2016patch,lebrun2013nonlocal,FFDNet,tai2017memnet,vaksman2020lidia,UDNet,zhang2020residual}.
Nowadays, supervised deep learning-based schemes lead the image denoising field, showing state-of-the-art (SoTA) performance~\cite{DnCNN,liu2018non,zhang2020residual}.

Common to these and many other algorithms is the fact that they minimize the expected distance, most notably the $L_2$ metric, between the original and the reconstructed images. This approach leads to a minimum mean squared error (MMSE) estimator. Unfortunately, high performance in terms of MSE does not necessarily mean good perceptual quality~\cite{wang2009mean}. Since denoising is an ill-posed problem (\ie a given input may have multiple correct solutions), the MMSE solutions tend to average these possible correct outcomes. In a high noise scenario, such an averaging strategy often leads to output images with blurry edges and unclear fine details. Many alternative distance measures to the MSE have been suggested, including SSIM~\cite{wang2004image}, MS-SSIM~\cite{wang2003multiscale}, IFC~\cite{sheikh2005information}, VIF~\cite{sheikh2006image}, VSNR~\cite{chandler2007vsnr}, and FSIM~\cite{zhang2011fsim}. However, changing the distance measure might not solve the problem. The authors of~\cite{blau2018perception, blau2018perception__CVF} have shown that there is an inherent contradiction between any mean distortion measure and perceptual quality. Due to this so-called "perception-distortion" trade-off, an image that minimizes the mean distance in any metric will necessarily suffer from a degradation in perceptual quality.

What could be the remedy to the above-described problem? While maintaining the Bayesian point of view, denoising could still leverage the posterior distribution of the unknown image given the measurements, but avoid the averaging effect. Seeking the highest peak of the posterior distribution or sampling from it, both seem as good strategies for getting high-perceptual quality outcomes.
\setcounter{footnote}{-1}
\begin{figure}
    \centering
    \includegraphics[width=\linewidth]{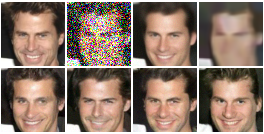}
    \caption[caption]{Top row, left to right: original image, noisy image with $\sigma = 0.406$ (pixel values are in the range $[0,1]$), the denoised results using an MMSE denoiser,\footnotemark~and BM3D~\cite{dabov2007image}. Bottom row: several outputs of our algorithm using the same MMSE denoiser.}
    \label{fig:flagship}
\end{figure}
\setcounter{footnote}{1}
\footnotetext{The MMSE denoiser used in all our experiments is based on NCSNv2~\cite{song2020improved}. See \autoref{sec:mmse_score} for more details.}

And indeed, many model-based classically-oriented algorithms seem to have chosen to apply a maximum a posteriori (MAP) estimator instead of MMSE (\eg ~\cite{elad2006image, zoran2011learning}).
MAP or closely related prior-based approaches are also applied more recently in deep learning based methods~\cite{ulyanov2018dip, golts2018deep, bhanumathi2017image, zhou2019towards}. However, these methods do not attempt to recover a highly probable natural-looking solution to the problem, but rather attempt to leverage a prior assumption on the image distribution in order to improve MSE performance. This is evident in the fact that the main evaluation metric for these methods is almost always peak signal to noise ratio (PSNR). In fact, many of these methods incorporate certain techniques in addition to MAP estimation, such as early stopping, patch averaging, or extra regularization, all done in order to achieve better PSNR performance, getting as close as possible to MMSE performance.

An alternative strategy to MAP is a sampler from the posterior. Recently, generative adversarial networks (GANs) have achieved success in generating realistically looking images (\eg~\cite{brock2018large,karras2020analyzing}), effectively sampling from the distribution of images. A GAN can serve our denoising task in one of two ways: either being trained to sample from the posterior directly, or by inverting its pre-trained generator. The first path refers to a variant of a conditional GAN, an approach that encounters difficulties in training, as exposed in~\cite{adler2018deep, adler2019deep}. The inversion option is appealing~\cite{lei2019inverting, aberdam2020}, but relies on an adversarial training, which is usually unstable, and there are currently no theoretical guarantees that its results are valid samples from the posterior distribution.

In this paper, we take a completely different approach towards high perceptual quality denoising that does not rely on GANs.
We draw inspiration from an interesting line of work reported in~\cite{song2019generative, song2020improved, ho2020denoising, song2020score} that develops an alternative method for generating images. The authors of~\cite{song2019generative} propose an \textit{annealed Langevin dynamics} algorithm in order to sample from the prior image distribution. This requires knowledge of the \textit{(Stein) score function}, which is the gradient of the log of the prior. The work in~\cite{kadkhodaie2020solving} introduces an extremely valuable link between this score function and MMSE denoisers, showing how to synthesize images and solve a special class of inverse problems by leveraging a given MMSE denoiser.

In our work we propose a novel approach for handling the image denoising task by building on the above. Our method produces sharp output images bypassing the typical denoisers' blurriness problem. Instead of minimizing MSE, our \textit{stochastic denoiser} samples its output from the posterior distribution given a corrupted image. The proposed algorithm stochastically picks a clean image consistent with the corrupted input one, instead of producing a single averaged output.
Similar ideas arise in recently published papers~\cite{bahat2020explorable,menon2020pulse}, which suggest that methods solving super resolution should not be deterministic, but rather allow for many possible outcomes.
In addition, stochastic denoising has been suggested in~\cite{wong2011stochastic}, but their method utilizes a far more complicated posterior sampling, and they use it for estimating an MMSE denoiser. In contrast, our algorithm samples directly from the posterior, achieving better perceptual quality.

For implementing sampling from the posterior, we formulate a score function that corresponds to the posterior, employ an MMSE denoiser to assist in evaluating it, and harness the annealed Langevin dynamics for drawing samples from this distribution. For any noisy input image, the proposed algorithm can produce a variety of viable outputs. Besides being sharp and natural-looking, images produced by the proposed stochastic denoiser are close to the MMSE solution in terms of PSNR, and visually similar to the true clean image. In fact, our work shows that all reconstructed images are valid outcomes of the denoising procedure, \ie the difference between any pair of noisy and reconstructed images is statistically fitted to an additive white Gaussian noise with the appropriate variance.

In addition, we introduce an extension of the stochastic denoising scheme for solving the noisy inpainting problem, in which the observed image is incomplete and contaminated by noise.
As in the denoising case, the inpainting scheme can produce a variety of valid yet different outputs for any input image.

Instead of working with a specific model architecture, our denoising and inpainting schemes utilize any denoiser trained/designed for minimizing the MSE on a set of noise levels. Such high-performance denoisers are widely available due to the incredible advances in image denoising achieved in the past two decades\footnote{Indeed, the extremely well-performing MMSE denoisers available today have led researchers to question whether we are nearing the optimal achievable noise reduction~\cite{chatterjee2009denoising, levin2012patch}.}.
Thus, our recovery schemes do not require any specific constraints on the model architecture or retraining of the MMSE denoiser. The only requirement is the ability to produce high-PSNR outputs for a set of noise levels.
To summarize, this paper has two main contributions:
\begin{itemize}
    \item We introduce a novel stochastic approach for the image denoising problem that leads to sharp and natural-looking reconstructions. Instead of reducing the restoration error, we propose to pick a probable solution by effectively sampling from the posterior distribution.
    \item We present stochastic algorithms for solving both the image denoising and inpainting problems.  For any corrupted input, these algorithms can produce a wide range of outputs where each is a possible valid solution of the problem.
\end{itemize}
\section{Proposed Method: Foundations}
\subsection{Sampling from the prior distribution}
One way to generate samples from a probability distribution $p\left(x\right)$, is using the Markov Chain Monte Carlo (MCMC) method with the Langevin transition rule~\cite{besag2001markov,roberts1996exponential}
\begin{equation}
\label{eqn:langevin_dynamics}
    x_{t+1} = x_t + \alpha \nabla_{x} \log p\left(x_t\right) + \sqrt{2\alpha} z_t \;,
\end{equation}
where $z_t \thicksim \mathcal{N}\left(0, I\right)$ and $\alpha$ some appropriate small constant. The expression $\nabla_{x} \log p\left(x\right)$ is known as the \emph{score function}~\cite{song2019generative} and is usually denoted as $s\left(x\right)$. The role of $z_t$ is to allow stochastic sampling, avoiding a collapse to a maximum of the distribution. Initialized randomly, after a sufficiently large number of iterations, and under some conditions, this process converges to a sampling from the desired distribution $p\left(x\right)$~\cite{roberts1996exponential}. The work reported in~\cite{song2019generative} extends the aforementioned algorithm to \textit{annealed Langevin dynamics}, which is handy for generating images from the implied prior distribution $p\left(x\right)$. The algorithm proposed by~\cite{song2019generative} works as follows: Initialized with a random image, it follows the direction of the score function in each step, as in \autoref{eqn:langevin_dynamics}. The score function is defined as $\nabla_{\tilde{x}} \log p\left(\tilde{x}\right)$ where $\tilde{x} = x + z$ and $z \thicksim \mathcal{N}\left(0, \sigma^{2}I\right)$ for different values of $\sigma$. Their method starts by using score functions corresponding to a high $\sigma$, and gradually lowers it until $p\left(\tilde{x}\right)$ is indistinguishable from the true data prior $p\left(x\right)$. This way, the algorithm flows the initial random image to ones with a higher prior probability, meaning that the output is a natural-looking image.

\subsection{Sampling from the posterior distribution}
We start by formulating our denoising task: Given a noisy input image $y = x + n$ where $x \thicksim p\left(x\right)$ is the true clean image and $n \thicksim \mathcal{N}\left(0, \sigma_{0}^{2}I\right)$ is a random white Gaussian noise with a known variance, we attempt to recover $x$.\footnote{Throughout this work we consider a Gaussian noise corruption, which provides a good approximation for many use cases~\cite{boyat2015}.}
This task might have multiple possible solutions for $x$, and we would like to output one of them. While $p\left(x\right)$ is unknown, we assume that we have access to an MMSE denoiser operating on images from $p\left(x\right)$. We propose to recover $x$ by sampling from the posterior distribution given the noisy input image, \ie, $p\left(x \mid y\right)$.

Our proposed approach is an adaptation of the \textit{annealed Langevin dynamics} algorithm~\cite{song2019generative} for our aforementioned task. \textit{Annealed Langevin dynamics} produces samples from $p\left(x\right)$ by means of the score function $\nabla_{\tilde{x}} \log p\left(\tilde{x}\right)$ where $\tilde{x} = x + z$ and $z \thicksim \mathcal{N}\left(0, \sigma^{2}I\right)$ for different values of $\sigma$. In order to adapt it to our task, we need to estimate the score function of the posterior $\nabla_{\tilde{x}} \log p\left(\tilde{x} \mid y\right)$. Note that the work reported in~\cite{kadkhodaie2020solving, song2020score} formulates score functions for posteriors for several inverse problems, and samples from these posteriors using Langevin dynamics. However, the problems treated are limited to noise free cases.

In the following we derive a formula for obtaining $\nabla_{\tilde{x}} \log p\left(\tilde{x} \mid y\right)$, and then present the relation between this score function and the MMSE denoiser. In \autoref{sec:algorithms} we present the stochastic denoiser algorithm for sampling from $p\left(x \mid y\right)$, which is based on the \textit{annealed Langevin dynamics} algorithm~\cite{song2019generative}. \renewcommand{\sectionautorefname}{Section}\autoref{sec:algorithms} also includes an extension of our algorithm for handling the inpainting problem.\renewcommand{\sectionautorefname}{section}
\begin{figure}
    \centering
    \includegraphics[width=0.7\linewidth]{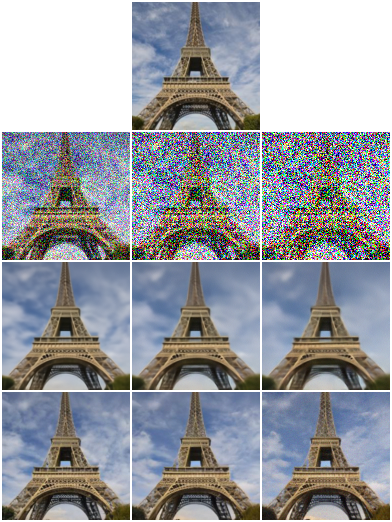}
    \caption{From top to bottom: original LSUN-tower image, noisy versions ($\sigma_0$ from left to right: $0.198$, $0.403$, $0.606$), MMSE denoiser outputs, and instances of our algorithm's output.}
    \label{fig:comparison}
\end{figure}

\subsubsection{The score function of the posterior distribution}
\label{sec:score_calc}

We fix a sequence of noise levels $\left\{\sigma_{i}\right\}_{i=0}^{L+1}$ such that ${\sigma_0 >\sigma_1 > ~~\cdots~~>\sigma_L>\sigma_{L+1}=0}$, where $\sigma_0$ is the noise level in $y$ and $\sigma_{L+1}$ is simply zero. We consider the process of adding white Gaussian noise with standard deviation $\sigma_0$ to $x$ as a gradual sequence of noise additions $\left\{\tilde{x}_i\right\}_{i=1}^{L}$ starting from $\tilde{x}_L$ down to $\tilde{x}_0$:
\begin{eqnarray}
\nonumber
{\tilde{x}}_{L} & = & x +z_{L}\\ 
\nonumber
{\tilde x}_{L-1} & = & {\tilde x}_{L} +z_{L-1}\\ 
\nonumber
{\tilde x}_{L-2} & = & {\tilde x}_{L-1} +z_{L-2}\\
& \vdots & \\
\nonumber {\tilde x}_{1} & = & {\tilde x}_{2} +z_{1}\\ 
\nonumber y = {\tilde x}_{0} & = & {\tilde x}_{1} +z_{0}
\end{eqnarray}
where $z_i \thicksim \mathcal{N}\left( 0,\left(\sigma_i^2-\sigma_{i+1}^2\right)I\right) $ for $0\le i \le L$
. Note that from the above we conclude that
\begin{equation}
    \label{eqn:y_def}
    y = \tilde{x}_{0} = x + \sum_{i=0}^{L} z_i,
\end{equation}
where $\sum_{i=0}^{L} z_i \thicksim \mathcal{N}\left(0, \sigma_{0}^{2}I\right)$, because a sum of independent Gaussian random variables is a Gaussian random variable with variance equal to the sum of their variances. This matches our original definition of $y$ in the denoising task. We also notice that
\begin{equation}
    \label{eqn:gauss_diff}
    y - \tilde{x}_{i} = \sum_{j=0}^{i-1} z_j,
\end{equation}
where $\sum_{j=0}^{i-1} z_j \thicksim \mathcal{N}\left(0, \left(\sigma_{0}^{2} - \sigma_{i}^{2}\right)I\right)$.

In the next calculations, which are valid for every $i$, we refer to $\tilde{x}_i$ as $\tilde{x}$ for simplicity. We move on to calculate $\nabla_{\tilde{x}} \log p\left(\tilde{x} \mid y\right)$ using the Bayes rule,
\begin{equation*}
\begin{aligned}
    \nabla_{\tilde{x}} \log p\left(\tilde{x} \mid y\right) = 
    \nabla_{\tilde{x}} \log \left[ \left(\frac{1}{p\left(y\right)}\right) p\left(y \mid \tilde{x}\right) p\left(\tilde{x}\right) \right] \\
    \hspace{0.1in}= \nabla_{\tilde{x}} \left[ \log \left(\frac{1}{p\left(y\right)}\right) + \log p\left(y \mid \tilde{x}\right) + \log p\left(\tilde{x}\right) \right].
\end{aligned}
\end{equation*}
Since $y$ is a fixed observation that does not depend on $\tilde{x}$, the gradient of the first term vanishes, resulting in
\begin{equation}
    \label{eqn:bayes}
    \nabla_{\tilde{x}} \log p\left(\tilde{x} \mid y\right) =
    \nabla_{\tilde{x}} \log p\left(y \mid \tilde{x}\right) + \nabla_{\tilde{x}} \log p\left(\tilde{x}\right).
\end{equation}

\noindent In order to calculate $\nabla_{\tilde{x}} \log p\left(y \mid \tilde{x}\right)$, we recall \autoref{eqn:gauss_diff} and obtain
\begin{equation*}
\begin{aligned}
    & \nabla_{\tilde{x}} \log p\left(y \mid \tilde{x}\right) =
    \nabla_{\tilde{x}} \log p_{Y - \tilde{X}}\left(y - \tilde{x} \mid \tilde{x}\right) \\
    & \hspace{0.1in} = \nabla_{\tilde{x}} \log \left[ \frac{1}{\sqrt{2\pi\left(\sigma_{0}^{2} - \sigma_{i}^{2}\right)}} \exp \left[ {- \frac{1}{2} \frac{\left\lVert y - \tilde{x}\right\rVert^2}{\left(\sigma_{0}^{2} - \sigma_{i}^{2}\right)}} \right] \right].
\end{aligned}
\end{equation*}
Calculating this results in
\begin{equation}
   \nabla_{\tilde{x}} \log p\left(y \mid \tilde{x}\right) =
   \frac{y - \tilde{x}}{\sigma_{0}^{2} - \sigma_{i}^{2}},
\end{equation}
which when combined with \autoref{eqn:bayes} gives
\begin{equation}
    \label{eqn:posterior_score}
    \nabla_{\tilde{x}} \log p\left(\tilde{x} \mid y\right) =
    \nabla_{\tilde{x}} \log p\left(\tilde{x}\right) + \frac{y - \tilde{x}}{\sigma_{0}^{2} - \sigma_{i}^{2}}.
\end{equation}
The first term is the same one used in~\cite{song2019generative}, while the second can be computed easily. Therefore, we have obtained a tractable method for estimating the score function of the posterior distribution given the noisy image.

\subsubsection{Estimating the score using an MMSE denoiser}
\label{sec:mmse_score}
A major step forward is provided in~\cite{kadkhodaie2020solving}, exposing the following intricate and fascinating connection between the score function and MMSE denoisers:
\begin{equation}
    \nabla_{\tilde{x}} \log p\left(\tilde{x}\right) =
    \frac{\hat{x}\left(\tilde{x}\right) - \tilde{x}}{\sigma_{i}^{2}},
\end{equation}
where $\hat{x}\left(\tilde{x}\right) = \mathbb{E}\left[x \mid \tilde{x}\right]$ is defined as the MMSE denoiser. This relation suggests that a network trained to estimate the score function (NCSNv2~\cite{song2020improved}) can be interpreted as a denoiser estimating MMSE. Indeed, we have utilized this network as such, and it performs very well, as can be seen in the MMSE denoiser results presented in \autoref{fig:flagship} and \autoref{tab:psnr}.

Likewise, we can utilize in our scheme any denoiser trained/designed to minimize MSE (for various noise levels $\sigma$) in order to estimate the score function.
A variety of such denoisers exist, each implicitly defining and serving a different prior distribution.
Adopting a broader view, the fact that MMSE denoisers can be leveraged for different tasks is a fascinating phenomenon, which has already been exposed in recent work in different contexts~\cite{plugNplay, romano2017red, kadkhodaie2020solving}.

\section{Proposed Algorithms}
\label{sec:algorithms}
\subsection{Stochastic denoising}
\label{sec:langevin_denoising}
In order to clean a given noisy image $y$, we propose to gradually reverse the noise addition process described in \autoref{sec:score_calc}. We do so stochastically, using a variation on the \textit{annealed Langevin dynamics}~\cite{song2019generative} sampling algorithm. We denote by $s\left(x, \sigma\right)$ a function that estimates the score function of the prior, and we present our method in \autoref{alg:sid}.
The algorithm follows the direction of the conditional score function, with a step size of $\alpha_i$ that is gradually tuned down as the noise level decreases (see~\cite{song2019generative}).
\begin{figure}[!t]
\removelatexerror
\begin{algorithm}[H]
\label{alg:sid}
 \KwIn{$\left\{\sigma_{i}\right\}_{i=1}^{L}$, $\epsilon$, $T$, $y$, $\sigma_0$}
 Initialize $x_0 \leftarrow y$\\
 \For{$i$ $\leftarrow$ $1$ to $L$}{
   $\alpha_{i} \leftarrow \epsilon \cdot \sigma_{i}^{2}/\sigma_{L}^{2}$ \\
   \For{$t$ $\leftarrow$ $1$ to $T$}{
     Draw $z_{t} \thicksim \mathcal{N}\left(0, I\right)$ \\
     $\Delta_{t} \leftarrow s\left(x_{t-1}, \sigma_{i}\right) + \left({y - x_{t-1}}\right)/\left({\sigma_{0}^{2} - \sigma_{i}^{2}}\right)$ \\
     $x_{t} \leftarrow x_{t-1} + \alpha_{i} \Delta_{t} + \sqrt{2 \alpha_{i}} z_{t}$
   }
 $x_0 \leftarrow x_T$
 }
 \KwOut{$x_T$}
 \caption{Stochastic image denoiser}
\end{algorithm}
\end{figure}

Our algorithm is initialized with the given noisy image $y$, and it follows the \textit{annealed Langevin dynamics} scheme using the score function of the posterior as presented in \autoref{eqn:posterior_score}.
This allows it to effectively sample from the posterior distribution $p\left(\tilde{x}_{L} \mid y\right) \approx p\left(x \mid y\right)$, and thus be considered a \textit{stochastic image denoiser}.

\begin{figure}
    \centering
    \includegraphics[width=0.8\linewidth]{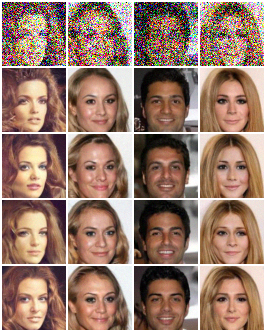}
    \caption{Top row: Several given noisy images ($\sigma_0 = 0.406$). Second row and below: Various outputs of our algorithm corresponding to each noisy image.}
    \label{fig:variation}
\end{figure}

\subsection{Image inpainting}
\label{sec:langevin_inpainting}
In the noisy inpainting problem, our observation is only a known subset $M$ of the pixels of the noisy image $y = x + n$. We denote the pixels $M$ of any image $w$ as $w^{M}$ and the remaining pixels as $w^{R}$. With these notations, the visible observation is $y^{M}$. However, our approach remains the same as in the denoising problem, aiming to sample from the posterior distribution $p\left(x \mid y^{M}\right)$.

As our observation is incomplete, 
we initialize the recovery algorithm with an i.i.d. Gaussian noise image with a very strong variance 
(as in~\cite{song2020improved}), and proceed from there by sampling from the posterior distribution given $y^{M}$.
More formally, we use a fixed sequence of noise levels $\left\{\sigma_i\right\}_{i=-K}^{L+1}$ such that ${\sigma_{-K}>\sigma_{-(K-1)}>~\cdots~>\sigma_0>~\cdots~>\sigma_L>\sigma_{L+1}=0}$, where 
$\sigma_0$ is the noise level of the observation.
In calculating the score function $\nabla_{\tilde{x}} \log p\left(\tilde{x} \mid y^{M}\right)$, we divide our analysis into two cases, $i<0$ in which the noise we handle is stronger than $\sigma_0$, and $i>0$, in which the noise bypasses $\sigma_0$ and gradually decreases towards zero. We start our derivations with the second case, as it is simpler: 

\begin{figure}
    \centering
    \includegraphics[width=0.75\linewidth]{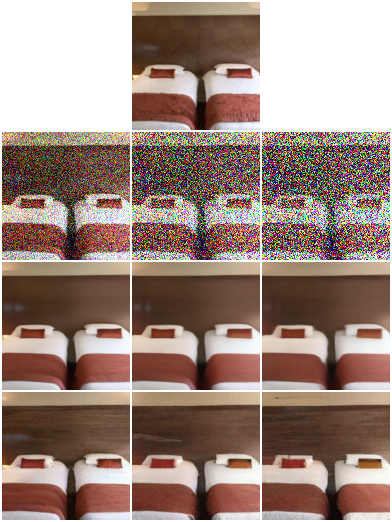}
    \caption{From top to bottom: original LSUN-bedroom image, noisy versions of it ($\sigma_0$ from left to right: $0.198$, $0.403$, $0.606$), MMSE denoiser outputs, and instances of our algorithm's output.}
    \label{fig:bedroom}
\end{figure}
\textbf{For the case where $\boldsymbol{i > 0}$}, we recall \autoref{eqn:gauss_diff} and deduce that
\begin{equation*}
\begin{aligned}
    & \nabla_{\tilde{x}} \log p\left(y^{M} \mid \tilde{x}\right) = \nabla_{\tilde{x}} \log p_{Y^{M} - \tilde{X}^{M}}\left(y^{M} - \tilde{x}^{M} \mid \tilde{x}\right) \\
    & \hspace{0.1in} = \nabla_{\tilde{x}} \log \left[ \frac{1}{\sqrt{2\pi\left(\sigma_{0}^{2} - \sigma_{i}^{2}\right)}} \exp \left[ {- \frac{1}{2} \frac{\left\lVert y^{M} - \tilde{x}^{M}\right\rVert^2}{\left(\sigma_{0}^{2} - \sigma_{i}^{2}\right)}} \right] \right].
\end{aligned}
\end{equation*}
Calculating this results in
\begin{equation}
   \left\{ \begin{array}{ll}
   \nabla_{\tilde{x}^{M}} \log p\left(y^{M} \mid \tilde{x}\right) =
   \frac{y^{M} - \tilde{x}^{M}}{\sigma_{0}^{2} - \sigma_{i}^{2}} \\
   \nabla_{\tilde{x}^{R}} \log p\left(y^{M} \mid \tilde{x}\right) = 0
   \end{array} \right.,
\end{equation}
which when combined with \autoref{eqn:bayes} gives the score function to use:
\begin{equation}
   \label{eqn:inp_i_smaller}
   \left\{ \begin{array}{ll}
   \nabla_{\tilde{x}^{M}} \log p\left( \tilde{x} \mid y^{M} \right) =
   \left[ \nabla_{\tilde{x}} \log p\left(\tilde{x}\right) \right]^{M} + 
   \frac{y^{M} - \tilde{x}^{M}}{\sigma_{0}^{2} - \sigma_{i}^{2}} \\
   \nabla_{\tilde{x}^{R}} \log p\left( \tilde{x} \mid y^{M} \right) = 
   \left[ \nabla_{\tilde{x}} \log p\left(\tilde{x}\right) \right]^{R}.
   \end{array} \right.\hspace{-0.2in}
\end{equation}
Since the noise level in this case is below $\sigma_0$, we effectively obtain the same score function as in the denoising task for the observed pixels $M$. As for the remaining pixels, $R$, the observation does not add any information, leaving us to rely only on the prior distribution.

\textbf{For the other case where $\boldsymbol{i < 0}$}, we recall the definition of the conditional distribution,
\begin{equation*}
\begin{aligned}
    & p\left(\tilde{x}^{R} \mid \tilde{x}^{M}, y^{M}\right) =
    \frac{p\left(\tilde{x}^{M}, \tilde{x}^{R} \mid y^{M}\right)}{p\left(\tilde{x}^{M} \mid y^{M}\right)} \\
    \Rightarrow & p\left(\tilde{x}^{R} \mid \tilde{x}^{M}, y^{M}\right) p\left(\tilde{x}^{M} \mid y^{M}\right) =
    p\left(\tilde{x}^{M}, \tilde{x}^{R} \mid y^{M}\right)
\end{aligned}
\end{equation*}
With that in mind, we present the following calculation of the log of the posterior function:
\begin{equation}
\label{eqn:inp_posterior}
\begin{aligned}
    & \log p\left(\tilde{x} \mid y^{M}\right) = \log p\left(\tilde{x}^{M}, \tilde{x}^{R} \mid y^{M}\right) \\
    &\hspace{0.1in} = \log \left[ p\left(\tilde{x}^{R} \mid \tilde{x}^{M}, y^{M}\right) p\left(\tilde{x}^{M} \mid y^{M}\right) \right] \\
    &\hspace{0.1in} = \log p\left(\tilde{x}^{R} \mid \tilde{x}^{M}, y^{M}\right) + \log p\left(\tilde{x}^{M} \mid y^{M}\right).
\end{aligned}
\end{equation}
Referring to the second term, we can conclude that, similar to \autoref{eqn:gauss_diff}, we have ${\left(\tilde{x}^{M} - y^{M}\right) \thicksim \mathcal{N}\left(0, \left(\sigma_{i}^{2} - \sigma_{0}^{2}\right)I\right)}$.
This difference is independent of $\tilde{x}^{R}$, and thus, $p\left(\tilde{x}^{R} \mid \tilde{x}^{M}, y^{M}\right)$ can be expressed as either $p\left(\tilde{x}^{R} \mid \tilde{x}^{M}\right)$ or $p\left(\tilde{x}^{R} \mid y^{M}\right)$. \autoref{eqn:inp_posterior} can be derived w.r.t. $\tilde{x}^{M}$:
\begin{equation*}
\begin{aligned}
    & \nabla_{\tilde{x}^{M}} \log p\left(\tilde{x} \mid y^{M}\right) \\
    &\hspace{0.1in} = \nabla_{\tilde{x}^{M}} \left[ \log p\left(\tilde{x}^{R} \mid y^{M}\right) + \log p_{\tilde{X}^{M} - Y^{M}}\left(\tilde{x}^{M} - y^{M} \mid y^{M}\right) \right] \\
    &\hspace{0.1in} \approx \nabla_{\tilde{x}^{M}} \log p_{\tilde{X}^{M} - Y^{M}}\left(\tilde{x}^{M} - y^{M} \mid y^{M}\right).
\end{aligned}
\end{equation*}
More details on this approximation are in~\autoref{sec:approx}. This results in
\begin{equation}
\label{eqn:inp_i_bigger_1}
    \nabla_{\tilde{x}^{M}} \log p\left(\tilde{x} \mid y^{M}\right) = 
    \frac{y^{M} - \tilde{x}^{M}}{\sigma_{i}^{2} - \sigma_{0}^{2}}.
\end{equation}
Deriving \autoref{eqn:inp_posterior} w.r.t. $\tilde{x}^{R}$ yields:
\begin{equation*}
\begin{aligned}
    & \nabla_{\tilde{x}^{R}} \log p\left(\tilde{x} \mid y^{M}\right) \\
    &\hspace{0.1in} = \nabla_{\tilde{x}^{R}} \left[ \log p\left(\tilde{x}^{R} \mid \tilde{x}^{M}\right) + \log p\left(\tilde{x}^{M} \mid y^{M}\right) \right] \\
    &\hspace{0.1in} = \nabla_{\tilde{x}^{R}} \log p\left(\tilde{x}^{R} \mid \tilde{x}^{M}\right) = \nabla_{\tilde{x}^{R}} \log \left[ \frac{p\left(\tilde{x}^{R}, \tilde{x}^{M}\right)}{p\left(\tilde{x}^{M}\right)} \right] \\
    &\hspace{0.1in} = \nabla_{\tilde{x}^{R}} \left[ \log p\left(\tilde{x}\right) - \log p\left(\tilde{x}^{M}\right) \right] \\
    &\hspace{0.1in} = \nabla_{\tilde{x}^{R}} \log p\left(\tilde{x}\right) = \left[ \nabla_{\tilde{x}} \log p\left(\tilde{x}\right) \right]^{R},
\end{aligned}
\end{equation*}
resulting in
\begin{equation}
\label{eqn:inp_i_bigger_2}
    \nabla_{\tilde{x}^{R}} \log p\left(\tilde{x} \mid y^{M}\right) =
    \left[ \nabla_{\tilde{x}} \log p\left(\tilde{x}\right) \right]^{R}.
\end{equation}
As the noise level in this case is above $\sigma_0$, the score function for the known pixels $M$ points to the direction of the observation $y^M$, which can be considered a denoised version of $\tilde{x}^M$. For the remaining pixels, $R$, the score function remains the same as in the previous case.

\textbf{To conclude}, by using an estimator for $\nabla_{\tilde{x}} \log p\left(\tilde{x}\right)$ and combining equations \ref{eqn:inp_i_smaller}, \ref{eqn:inp_i_bigger_1}, and \ref{eqn:inp_i_bigger_2}, we obtain a tractable method for estimating the score function of the posterior distribution for the inpainting problem.
By using this and starting \autoref{alg:sid} with a very strong noise $\sigma_{-K}\gg\sigma_{0}$, we obtain a path towards solving the noisy inpainting problem, as presented in \autoref{alg:inp}.

\begin{table}[]
    \centering
    \begin{tabular}{|c|c|c c c|c|} 
        \hline
        \textbf{Dataset} & $\mathbf{\sigma_{0}}$ & \textbf{BM3D} & \textbf{MMSE} & \textbf{Ours} & \textbf{Ratio}\\
        \hline
        \multirow{5}{*}{CelebA}
        &$0.100$ & $30.18$ & ${32.58}$ & $29.39$ & $2.07$\\ 
        &$0.203$ & $25.43$ & ${29.08}$ & $26.28$ & $1.90$\\ 
        &$0.406$ & $19.73$ & ${25.78}$ & $23.24$ & $1.79$\\ 
        &$0.607$ & $16.75$ & ${23.93}$ & $21.52$ & $1.73$\\ 
        &$0.702$ & $15.87$ & ${23.27}$ & $20.90$ & $1.72$\\ 
        \hline
        \multirow{3}{*}{\makecell{LSUN\\bedroom}}
        &$0.198$ & $27.19$ & ${29.95}$ & $27.11$ & $1.91$\\ 
        &$0.403$ & $21.31$ & ${26.50}$ & $24.00$ & $1.78$\\ 
        &$0.606$ & $18.17$ & ${24.55}$ & $22.22$ & $1.72$\\ 
        \hline
    \end{tabular}
    \caption{Average PSNR results using 64 CelebA images and 64 LSUN images, including BM3D~\cite{dabov2007image} as a baseline. The last column shows the MSE ratio between the MMSE and ours.}
    \label{tab:psnr}
\end{table}

\begin{figure}[!t]
\removelatexerror
\begin{algorithm}[H]
\label{alg:inp}
 \KwIn{$\left\{\sigma_{i}\right\}_{i=-K}^{L}$, $\epsilon$, $T$, $y^M$}
 Initialize $x_0$ with random noise\\
 \For{$i$ $\leftarrow$ $-K$ to $-1$}{
   $\alpha_{i} \leftarrow \epsilon \cdot \sigma_{i}^{2}/\sigma_{L}^{2}$ \\
   \For{$t$ $\leftarrow$ $1$ to $T$}{
     Draw $z_{t} \thicksim \mathcal{N}\left(0, I\right)$ \\
     $\Delta_{t}^{M} \leftarrow \left({y^{M} - x_{t-1}^{M}}\right)/\left({\sigma_{i}^{2} - \sigma_{0}^{2}}\right)$ \\
     $\Delta_{t}^{R} \leftarrow \left[ s\left(x_{t-1}, \sigma_{i}\right) \right]^{R}$ \\
     $x_{t} \leftarrow x_{t-1} + \alpha_{i} \Delta_{t} + \sqrt{2 \alpha_{i}} z_{t}$
   }
 $x_0 \leftarrow x_T$
 }
 \For{$i$ $\leftarrow$ $1$ to $L$}{
   $\alpha_{i} \leftarrow \epsilon \cdot \sigma_{i}^{2}/\sigma_{L}^{2}$ \\
   \For{$t$ $\leftarrow$ $1$ to $T$}{
     Draw $z_{t} \thicksim \mathcal{N}\left(0, I\right)$ \\
     $\Delta_{t}^{M} \leftarrow \left[ s\left(x_{t-1}, \sigma_{i}\right) \right]^{M}$ \\
     $\Delta_{t}^{M} \leftarrow \Delta_{t}^{M} + \left({y^{M} - x_{t-1}^{M}}\right)/\left({\sigma_{0}^{2} - \sigma_{i}^{2}}\right)$ \\
     $\Delta_{t}^{R} \leftarrow \left[ s\left(x_{t-1}, \sigma_{i}\right) \right]^{R}$ \\
     $x_{t} \leftarrow x_{t-1} + \alpha_{i} \Delta_{t} + \sqrt{2 \alpha_{i}} z_{t}$
   }
 $x_0 \leftarrow x_T$
 }
 \KwOut{$x_T$}
 \caption{Inpainting algorithm}
\end{algorithm}
\end{figure}
\section{Experimental Results}
\subsection{Denoising experiments}
\label{sec:denoising_exp}

As our algorithm extends the work reported in~\cite{song2019generative} and~\cite{song2020improved}, it is natural to use their denoiser network, named the noise conditional score network version 2 (NCSNv2)~\cite{song2020improved}. As is customary in the image synthesis literature, this network was trained on a specific class of images rather than generic natural ones. Previous work~\cite{remez2017deep, remez2018class} have also shown that denoising can benefit from training on a narrow class of images.

We perform experiments using the NCSNv2 network trained separately on CelebA~\cite{liu2015deep}, FFHQ~\cite{karras2020analyzing}, and the bedroom and tower categories in LSUN~\cite{yu2015lsun}. CelebA images are center cropped to $140 \times 140$ pixels, then resized to $64 \times 64$ pixels, FFHQ images are resized to $256 \times 256$ pixels, and LSUN images are center cropped and resized to $128 \times 128$ pixels, exactly as in~\cite{song2020improved}. We do not change the hyperparameters reported in~\cite{song2020improved}, as shown in \autoref{sec:hyperparameters}.
For CelebA experiments we pick ${L=127, 166, 204, 226, 234}$ for ${\sigma_{0} = 0.100, 0.203, 0.406, 0.607, 0.702}$ respectively.
For FFHQ experiments we pick ${L=663, 816, 906}$ for ${\sigma_{0}=0.200, 0.400, 0.602}$ respectively.
For LSUN experiments we pick ${L=330, 408, 453}$ for ${\sigma_{0}=0.198, 0.403, 0.606}$ respectively.
We note that in each of our experiments we use the same pre-trained model for both the MMSE denoiser and in our algorithm.

\begin{figure}
    \centering
    \setlength\tabcolsep{1.5pt}
    \begin{tabular}{cc}
        \setlength\tabcolsep{0.5pt}
        \renewcommand{\arraystretch}{0.5}
        \begin{tabular}{ccc}
            \includegraphics[width=0.16\linewidth]{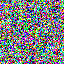}&
            \includegraphics[width=0.16\linewidth]{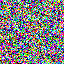}&
            \includegraphics[width=0.16\linewidth]{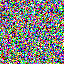}\\
            \scriptsize{(0.84, 0.02)} & \scriptsize{(0.82, 0.01)} & \scriptsize{(0.77, -0.01)}\\
            \\
            \includegraphics[width=0.16\linewidth]{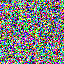}&
            \includegraphics[width=0.16\linewidth]{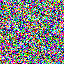}&
            \includegraphics[width=0.16\linewidth]{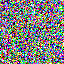}\\
            \scriptsize{(0.89, -0.02)} & \scriptsize{(0.82, -0.01)} & \scriptsize{(0.71, -0.01)}
        \end{tabular}
        &
        \begin{tabular}{c}
            \footnotesize{\ \ \ \ std=$0.60$, p-val=$0.82$, $\rho$=$0.01$} \\
            \includegraphics[width=0.44\linewidth]{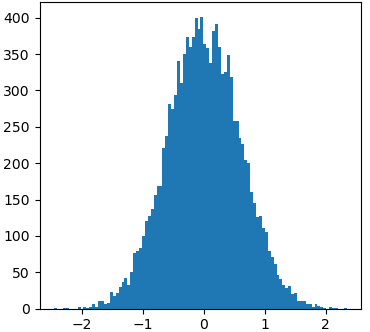}
        \end{tabular}
    \end{tabular}
    \caption{Left: Residual images, p-values and $\rho$ values on CelebA with ${\sigma_{0} = 0.607}$ for 3 different images. The standard deviation is $0.59$ or $0.6$ in all images. The top row shows our method's residuals, and the bottom one shows the MMSE denoiser's residuals. Right: A histogram for a residual image from our algorithm.}
    \label{fig:residuals}
\end{figure}

\begin{figure}
    \centering
    \includegraphics[width=0.76\linewidth]{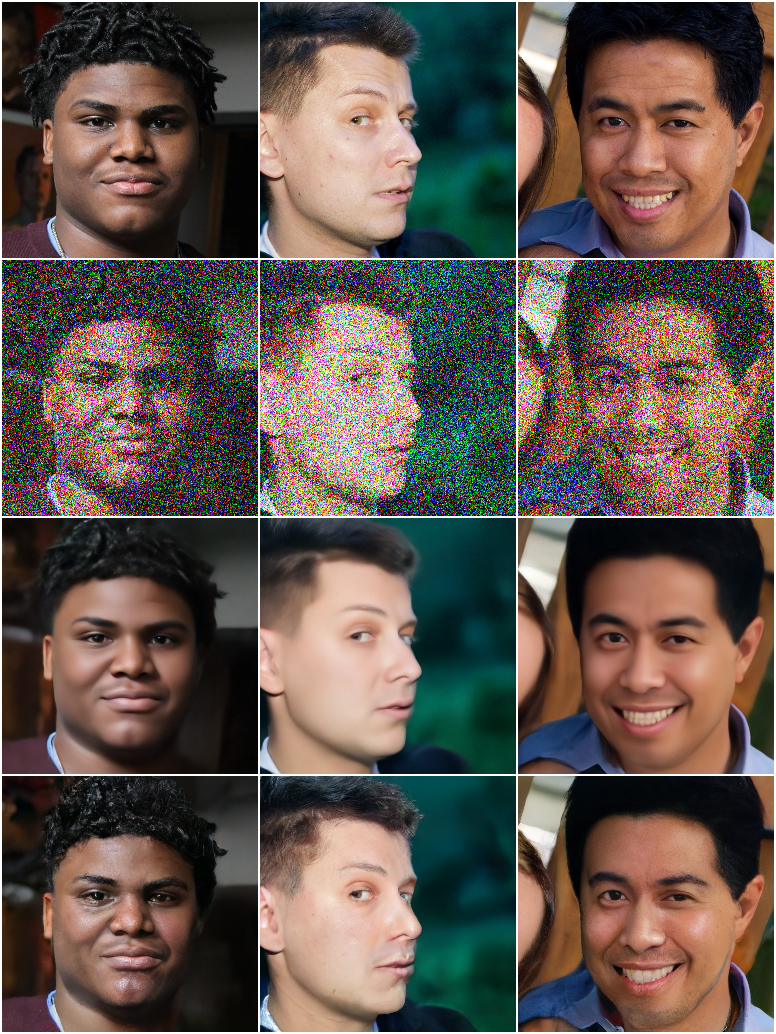}
    \caption{From top to bottom: original $256 \times 256$ FFHQ images, noisy versions with $\sigma_0 = 0.4$, and our algorithm's outputs.}
    \label{fig:ffhq}
\end{figure}

As can be seen in \autoref{fig:comparison}, \ref{fig:bedroom}, and \ref{fig:ffhq}, our stochatic denoising method achieves sharp and real-looking results, regardless of the noise level in the input image, whereas the MMSE denoiser suffers from more severe averaging artefacts as the noise level increases. The results' sharpness is also preserved across different stochastic variations, as can be seen in \autoref{fig:flagship}, \ref{fig:variation}, and in \autoref{sec:extra_figures}.

We evaluate the perceptual quality of the results using LPIPS~\cite{lpips}, in which our model performs significantly better than the MMSE denosier, as shown in~\autoref{sec:lpips}.
We also asses the similarity to the original clean image using the MSE metric (or its equivalent PSNR). While the MSE measure has clear drawbacks~\cite{wang2009mean}, and our algorithm inherently achieves poorer results than an MMSE denoiser, we still find value in reporting such results. It was proven in~\cite{blau2018perception, blau2018perception__CVF} that we do not need to sacrifice more than a factor of $2$ in MSE in order to achieve perfect perceptual quality, which serves as a good baseline for us to evaluate our results. As can be seen in \autoref{tab:psnr}, the aforementioned ratio is comfortably below $2$ in all experiments but one.

\subsection{Assesing the estimated noise}
We now turn to show that all outputs of the presented algorithm are viable denoising results. A sample from $p\left(x \mid y\right)$ should both look real and fulfill the condition ${\left(y - \hat{x}\right) \thicksim \mathcal{N}\left(0, \sigma_{0}^{2}I\right)}$. The latter is also a criterion for claiming that a given algorithm is a denoiser, as it suggests that the content removed from its input is indeed noise-like. MMSE denoisers, for example, fulfill this criterion.

In order to empirically test whether our algorithm is a stochastic image denoiser as we claim, we analyze the estimated noise -- the difference between its input and output, as visualized in \autoref{fig:residuals}.
Our analysis includes three tests: for whiteness, for noise energy, and for the distribution. First, we calculate Pearson’s correlation coefficient ($\rho$) among adjacent pixels in all 8 directions, take the one with the maximum absolute value, and if it is sufficiently close to zero, we conclude that the noise is uncorrelated. 
We proceed by performing D’Agostino and Pearson’s test of normality~\cite{dagostino} in order to determine whether the difference is normally distributed. For a confidence level of $95\%$, we conclude that the tested signal is indeed Gaussian if the p-value is greater than $0.05$. We conclude by evaluating the empirical standard deviation of the difference and comparing it to $\sigma_0$. 

\begin{figure*}
    \centering
    \includegraphics[width=\linewidth]{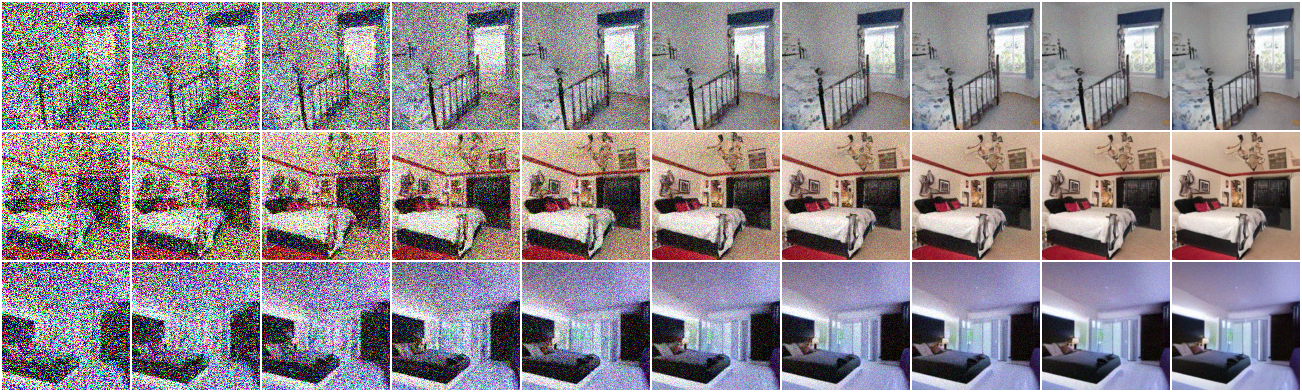}
    \caption{Intermediate results of \autoref{alg:sid} on LSUN-bedroom images with $\sigma_0=0.606$.}
    \label{fig:progression}
\end{figure*}

We perform these tests on several output images for each noisy input and different noise levels. In almost all of our tests, $\left|\rho\right|$ is smaller than $0.02$, the p-values are comfortably above $0.05$, often reaching more than $0.9$, and the standard deviations match the input noise level $\sigma_0$ almost perfectly. We show one of the residual histograms in \autoref{fig:residuals}, and more histograms in \autoref{sec:extra_figures}.
Based on these observations, we conclude that our sampled images can be regarded as viable stochastic denoising results.
\autoref{fig:progression} shows the intermediate images obtained along our algorithm, showing a gradual denoising effect, while preserving and even synthesizing details. 

\begin{figure}
    \centering
    \includegraphics[width=0.9\linewidth]{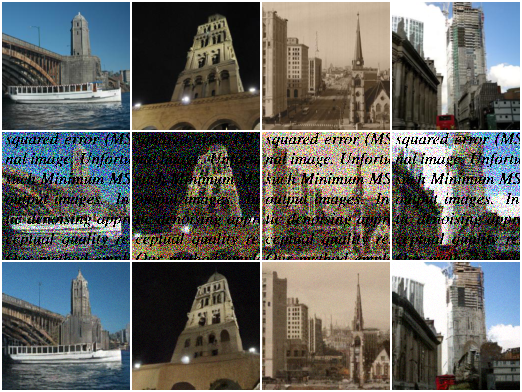}
    \caption{From top to bottom: original LSUN-tower images, the observations with a text overlay and additive noise ($\sigma_0 = 0.198$), and our inpainting algorithm's outputs.}
    \label{fig:inpainting_eye}
\end{figure}

\subsection{Inpainting experiments}
\begin{figure}
    \centering
    \includegraphics[width=0.77\linewidth]{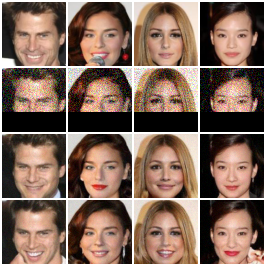}
    \caption{From top to bottom: original CelebA images, the observations with $20$ missing rows and additive noise ($\sigma_0 = 0.1$), and two outputs of our inpainting algorithm.}
    \label{fig:inpainting}
\end{figure}

Following the calculations shown in \autoref{sec:langevin_inpainting}, we adapt our stochastic denoiser algorithm for solving the noisy inpainting task. As in \autoref{sec:denoising_exp}, we utilize the NCSNv2~\cite{song2020improved} network for estimating the score function of the prior distribution, and we perform experiments on the CelebA~\cite{liu2015deep} and LSUN~\cite{yu2015lsun} datasets. Here as well we do not change the hyperparameters reported in~\cite{song2020improved}.
We show results of our algorithm in \autoref{fig:inpainting_eye}, \ref{fig:inpainting}, and \autoref{sec:extra_figures}.
\section{Conclusion}
In this work we present a new image denoising approach, which samples from the posterior distribution given the noisy image. We argue that in order to attain high perceptual quality, a denoising algorithm should be stochastic rather than deterministic, having multiple possible outcomes. We present a denoising algorithm along these lines, showcasing high-quality results. Our method relies on the annealed Langevin dynamics algorithm, requiring only an MMSE denoiser, without any additional retraining, constraints on the model architecture, nor extra model parameters. In addition, we extend our algorithm for handling the problem of noisy image inpainting.

Our algorithm
 takes a significant amount of time ($\sim2$ minutes for $8$ CelebA images) 
in order to guarantee a proper convergence to a valid sampling result. Means to speed-up this procedure are therefore necessary. Our future work will focus on speeding this method by multi-scale denoising~\cite{block2020fast}, deployment of denoisers of varying complexities, and other acceleration techniques.
Other future research directions we consider include (\rNum{1}) treating general content images using general purpose denoisers, and handling much larger images (our current solution operates on images of up to size $256\times256$ pixels), (\rNum{2}) assessing the image manifold as implicitly implied by varying denoisers; and (\rNum{3}) developing uncertainty measures for the denoising solution to expose and quantify the diversity of the possible solutions.

\section{Acknowledgement}
We would like to thank Mauricio Belbracio and Peyman Milanfar from Google Research for drawing our attention to the work in~\cite{kadkhodaie2020solving, song2019generative}, which inspired our work.

{\small
\bibliographystyle{ieee_fullname}
\bibliography{egbib}
}

\pagebreak
\begin{appendices}
\section{Experimental Details}
\label{sec:hyperparameters}
\autoref{tab:deno} and \autoref{tab:inp} present the hyperparameters used in our experiments. The hyperparameter $\sigma_L$ is set to $0.01$ in all experiments.
\begin{table}[H]
    \centering
    \begin{tabular}{|c|c c c|}
        \hline
        \textbf{Dataset} & $\mathbf{\epsilon}$ & $\mathbf{T}$ & $\mathbf{\frac{\sigma_{i+1}}{\sigma_{i}}}$ \\
        \hline
        CelebA~\cite{liu2015deep}
        &$3.3e-6$ & ${5}$ & $0.982$\\
        \hline
        LSUN~\cite{yu2015lsun}
        &$1.8e-6$ & ${3}$ & $0.991$\\
        \hline
        FFHQ~\cite{karras2020analyzing}
        &$1.8e-6$ & ${3}$ & $0.995$\\
        \hline
    \end{tabular}
    \caption{Hyperparameters for our denoising experiments. The last column is the geometric progression rate for $\left\{\sigma_{i}\right\}_{i=0}^{L+1}$.}
    \label{tab:deno}
\end{table}
\begin{table}[H]
    \centering
    \begin{tabular}{|c|c c c c c|}
        \hline
        \textbf{Dataset} & $\mathbf{\epsilon}$ & $\mathbf{T}$ & $\mathbf{L+K}$ & $\mathbf{\sigma_{-K}}$ & $\mathbf{\frac{\sigma_{i+1}}{\sigma_{i}}}$ \\
        \hline
        CelebA~\cite{liu2015deep}
        &$3.3e-6$ & ${5}$ & $500$ & ${90}$ & $0.982$\\
        \hline
        LSUN~\cite{yu2015lsun}
        &$1.8e-6$ & ${3}$ & $1086$ & ${190}$ & $0.991$\\
        \hline
        FFHQ~\cite{karras2020analyzing}
        &$1.8e-6$ & ${5}$ & $2311$ & ${348}$ & $0.995$\\
        \hline
    \end{tabular}
    \caption{Hyperparameters for our inpainting experiments. 
    }
    \label{tab:inp}
\end{table}

\section{Perceptual Quality Evaluation}
\label{sec:lpips}
We evaluate the perceptual quality of the results using LPIPS~\cite{lpips} (version $0.1$), a distance metric shown to perform better as a perceptual metric than PSNR or SSIM. We evaluated 64 sets of images, each set containing an original CelebA image, a denoised one using an MMSE denoiser, and our algorithm's output, utilizing the same MMSE denoiser.
Our model performs significantly better than the MMSE denoiser, achieving around $16\%$ lower (and thus, better) LPIPS perceptual distance from the original image, as can be seen in \autoref{tab:lpips}.

\section{Inpainting Approximate Derivation}
\label{sec:approx}
The following approximation is used in the derivation of Equation 12, referring to the inpainting algorithm:
\begin{equation}
\label{eqn:approx}
    \nabla_{\tilde{x}^M} \log p\left(\tilde{x}^R | \tilde{x}^M\right) = \nabla_{\tilde{x}^M} \log p\left(\tilde{x}^R | y^M\right) \approx 0.
\end{equation}
We support this approximation by making the following observation:
\begin{align}
    & \nabla_{\tilde{x}^M} \log p\left(\tilde{x}^R | \tilde{x}^M\right) = \nabla_{\tilde{x}^M} \log \frac{p\left(\tilde{x}^R, \tilde{x}^M\right)}{p\left(\tilde{x}^M\right)} \nonumber \\
    & = \nabla_{\tilde{x}^M} \log p\left(\tilde{x}\right) - \nabla_{\tilde{x}^M} \log p\left(\tilde{x}^M\right)\\
    & = \left(\nabla_{\tilde{x}} \log p\left(\tilde{x}\right)\right)^M - \nabla_{\tilde{x}^M} \log p\left(\tilde{x}^M\right) \nonumber\\
    & = \mathbb{E}\left[x | \tilde{x}\right]^M - \mathbb{E}\left[x^M | \tilde{x}^M\right] 
    = \mathbb{E}\left[x^M | \tilde{x}\right] - \mathbb{E}\left[x^M | \tilde{x}^M\right]. \nonumber
\end{align}
The second to last equality holds due to Equation 8 in the paper. We obtained a difference between two estimators of $x^M$, both of which depend on information from a noisy version of it, $\tilde{x}^M$, but the second one contains additional information about $\tilde{x}^R$. While this information may change the estimation, we assume its impact to be negligible, resulting in this difference to be approximately zero, thus obtaining \autoref{eqn:approx}.

\begin{table}
    \centering
    \begin{tabular}{|c|c|c c|c|}
        \hline
        \textbf{Dataset} & $\mathbf{\sigma_{0}}$ & \textbf{Ours} & \textbf{MMSE} & \textbf{Improvement}\\
        \hline
        \multirow{3}{*}{CelebA~\cite{liu2015deep}}
        &$0.203$ & $0.027$ & $0.032$ & $15.6\%$\\ 
        &$0.406$ & $0.049$ & $0.059$ & $16.9\%$\\ 
        &$0.607$ & $0.067$ & $0.081$ & $17.3\%$\\ 
        \hline
    \end{tabular}
    \caption{Average LPIPS distance using 64 CelebA images. The last column shows the percentage of improvement that our algorithms presents.}
    \label{tab:lpips}
\end{table}
\section{Additional Results}
\label{sec:extra_figures}
In the following, we present additional qualitative results not shown in the paper. They are best viewed digitally, and zoomed in. Standard deviations, p-values, and correlation coefficients were rounded to 2 decimal places.

\begin{figure*}
    \centering
    \includegraphics[width=0.5\linewidth]{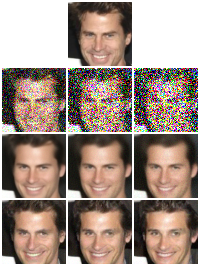}
    \caption{From top to bottom: original CelebA image, noisy versions ($\sigma_0$ from left to right: $0.203$, $0.406$, $0.607$), MMSE denoiser outputs, and instances of our denoising algorithm's output.}
    \label{fig:denoising_begin}
\end{figure*}

\begin{figure*}
    \centering
    \includegraphics[width=\linewidth]{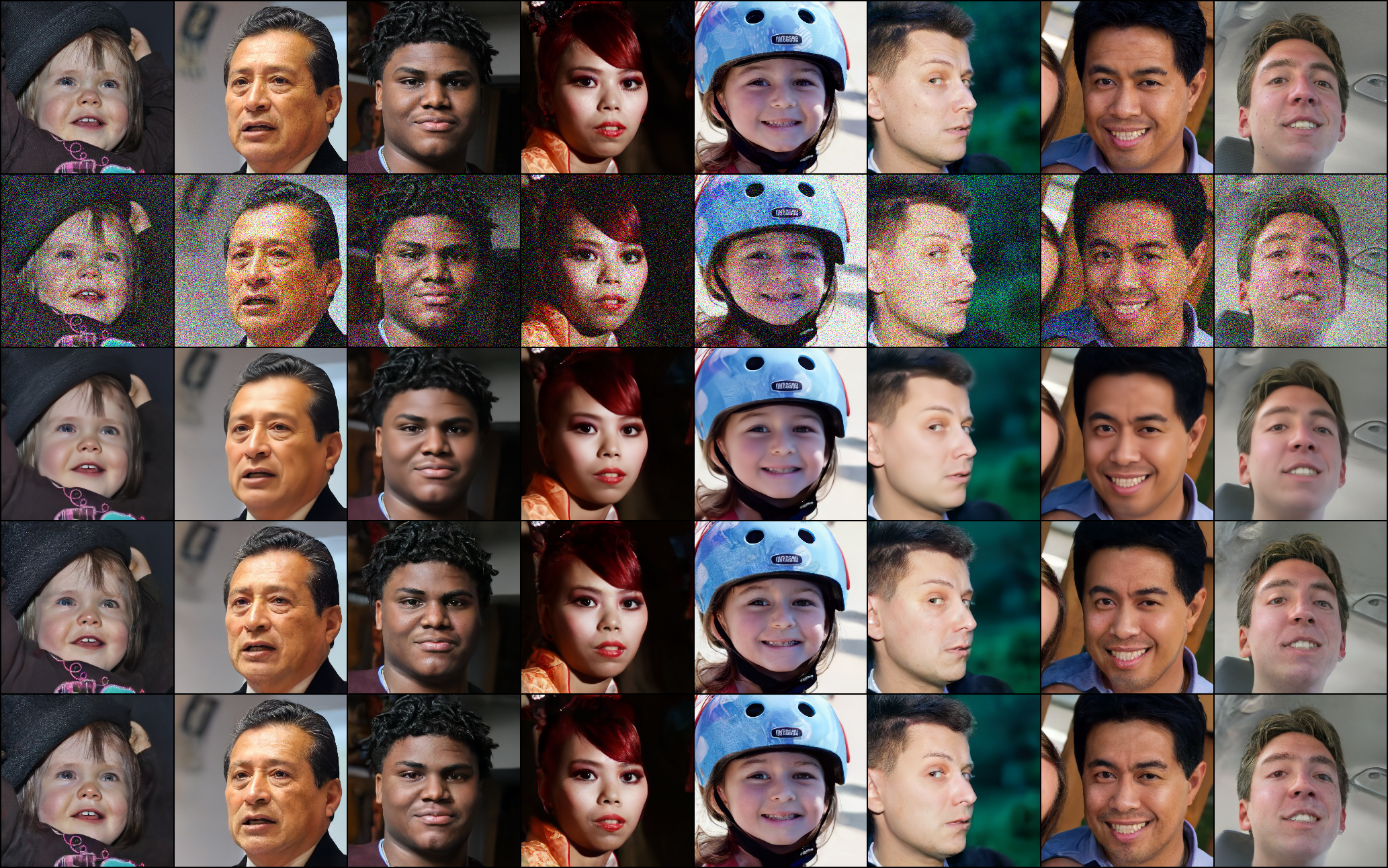}
    \caption{From top to bottom: original FFHQ images, noisy versions with $\sigma_0=0.2$, MMSE denoiser outputs, and two instances of our denoising algorithm's output.}
    \label{fig:denoising_begin2}
\end{figure*}

\begin{figure*}
    \centering
    \includegraphics[width=\linewidth]{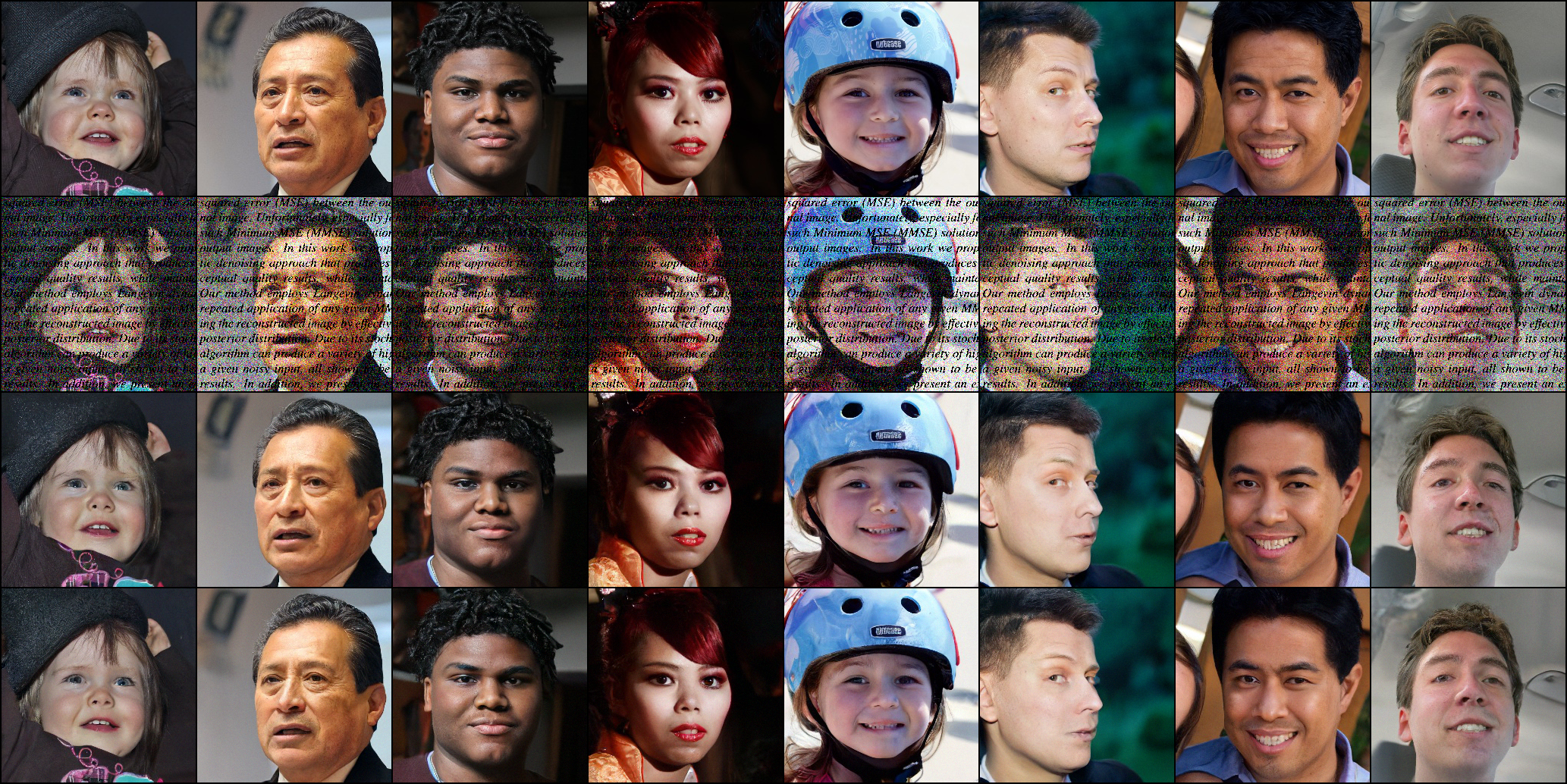}
    \caption{From top to bottom: original FFHQ images, the observations with text overlay and additive noise ($\sigma_0 = 0.2$), and two instances of our inpainting algorithm's output.}
    \label{fig:denoising_end}
\end{figure*}

Figures \ref{fig:denoising_begin}-\ref{fig:denoising_end} present additional material referring to our denoising scheme. \autoref{fig:denoising_begin} presents both MMSE and our denoising results for several noise levels, demonstrating the tendency of our method to produce far sharper images. Figures \ref{fig:denoising_begin2}-\ref{fig:denoising_end} show further denoising results for images taken from FFHQ and LSUN with varying noise levels, this time emphasizing the possible diversity of the outcomes of our stochastic denoiser, and their validity as denoising results.

\begin{figure*}
    \centering
    \includegraphics[width=\linewidth]{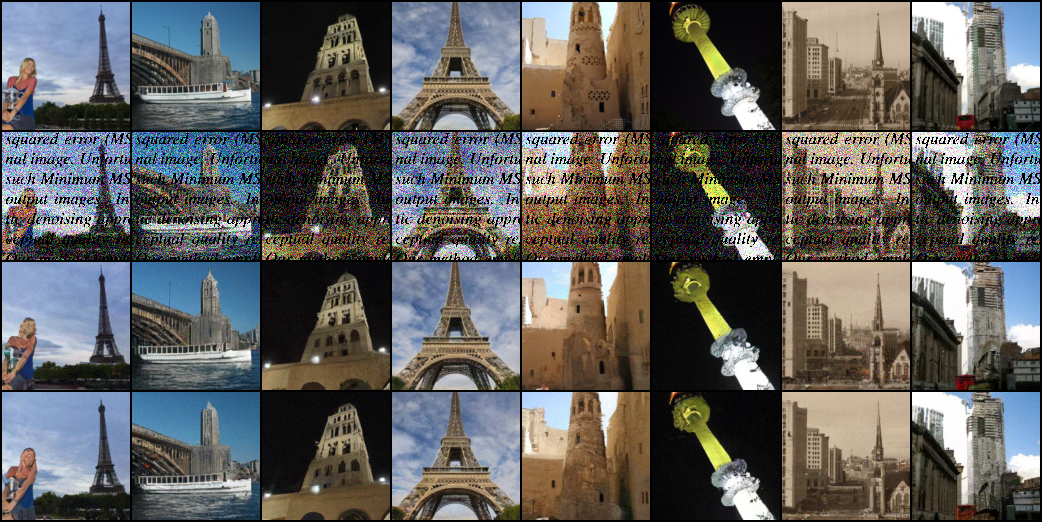}
    \caption{From top to bottom: original LSUN-tower images, the observations with text overlay and additive noise ($\sigma_0 = 0.198$), and two instances of our inpainting algorithm's output.}
    \label{fig:inpainting_begin}
\end{figure*}

\begin{figure*}
    \centering
    \includegraphics[width=0.5\linewidth]{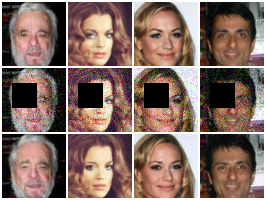}
    \caption{From top to bottom: original CelebA images, the observations with a missing eye and additive noise ($\sigma_0 = 0.1$), and outputs of our inpainting algorithm.}
    \label{fig:inpainting_end2}
\end{figure*}

\begin{figure*}
    \centering
    \includegraphics[width=\linewidth]{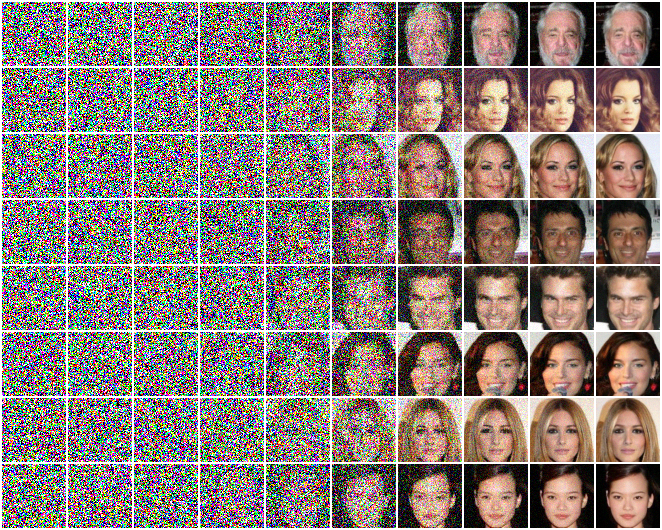}
    \caption{Intermediate results of our inpainting algorithm on CelebA images with a missing eye and additive noise of $\sigma_{0}=0.1$.}
    \label{fig:inpainting_end}
\end{figure*}

Figures \ref{fig:inpainting_begin}-\ref{fig:inpainting_end} introduce additional material referring to our inpainting scheme. Figures \ref{fig:inpainting_begin}-\ref{fig:inpainting_end2} showcase additional inpainting results, highlighting the sharpness of the results and the posterior distribution score function's ability to synthesize features consistent with the surroundings of the missing pixels. \autoref{fig:inpainting_end} shows intermediate results obtained along our inpainting algorithm for different images, gradually converging towards the final resulting images. 

\begin{figure*}
    \centering
    \includegraphics[width=\linewidth]{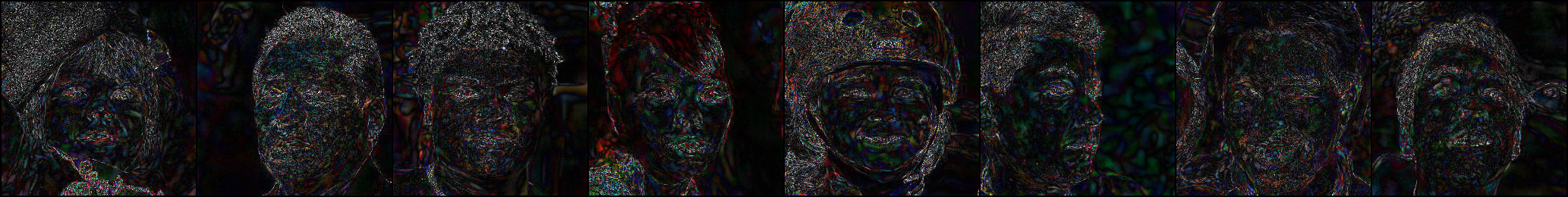}
    \caption{The absolute value of the difference (scaled by $4$) between two instances of our denoising algorithm's output on FFHQ images with $\sigma_0=0.2$.}
\end{figure*}
\begin{figure*}
    \centering
    \includegraphics[width=\linewidth]{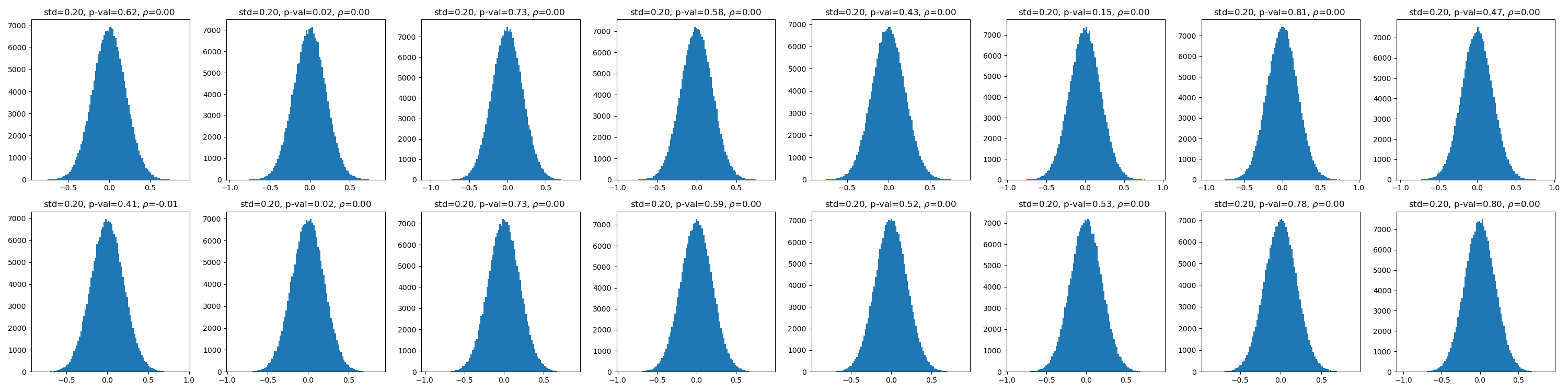}
    \caption{Residual histograms, standard deviations, normality p-values, and the Pearson's correlation coefficients (in the direction with the maximum absolute value) for our denoising algorithm's output on FFHQ images with $\sigma_0=0.2$.}
\end{figure*}

\begin{figure*}
    \centering
    \includegraphics[width=\linewidth]{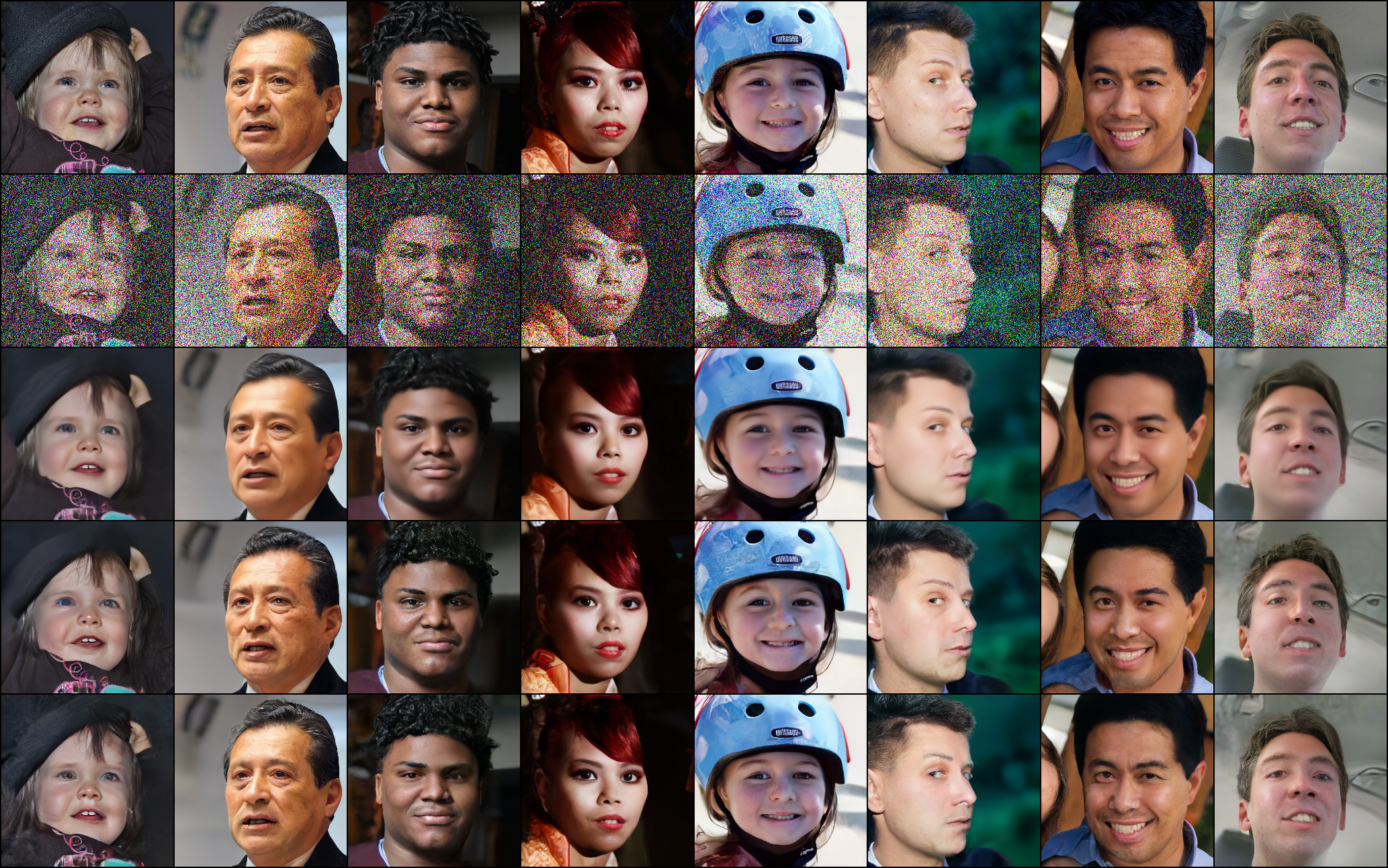}
    \caption{From top to bottom: original FFHQ images, noisy versions with $\sigma_0=0.4$, MMSE denoiser outputs, and two instances of our denoising algorithm's output.}
\end{figure*}
\begin{figure*}
    \centering
    \includegraphics[width=\linewidth]{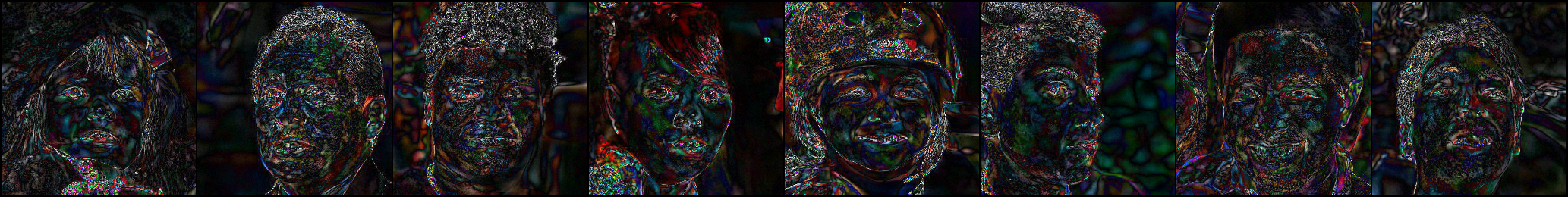}
    \caption{The absolute value of the difference (scaled by $4$) between two instances of our denoising algorithm's output on FFHQ images with $\sigma_0=0.4$.}
\end{figure*}
\begin{figure*}
    \centering
    \includegraphics[width=\linewidth]{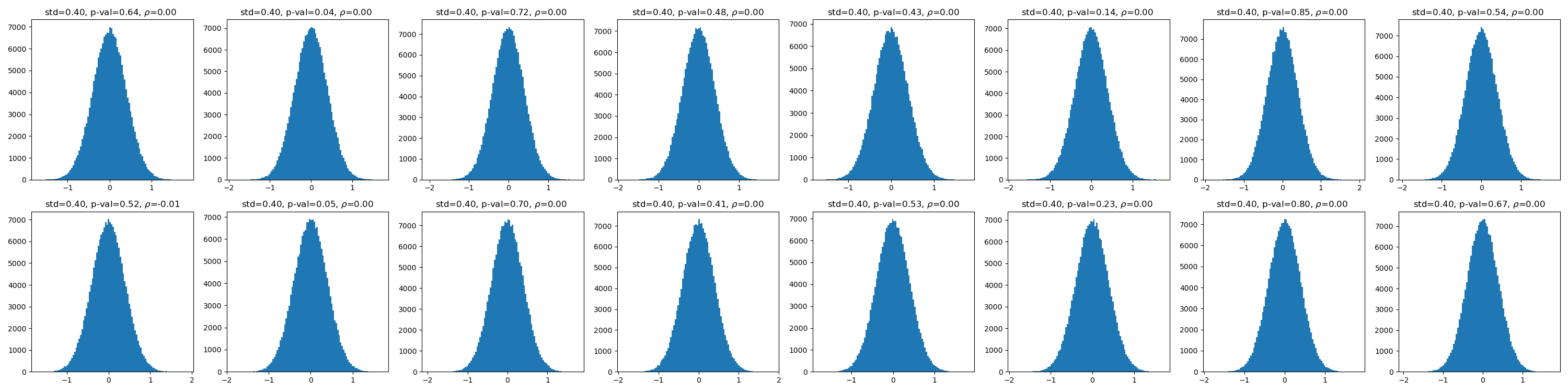}
    \caption{Residual histograms, standard deviations, normality p-values, and the Pearson's correlation coefficients (in the direction with the maximum absolute value) for our denoising algorithm's output on FFHQ images with $\sigma_0=0.4$.}
\end{figure*}

\begin{figure*}
    \centering
    \includegraphics[width=\linewidth]{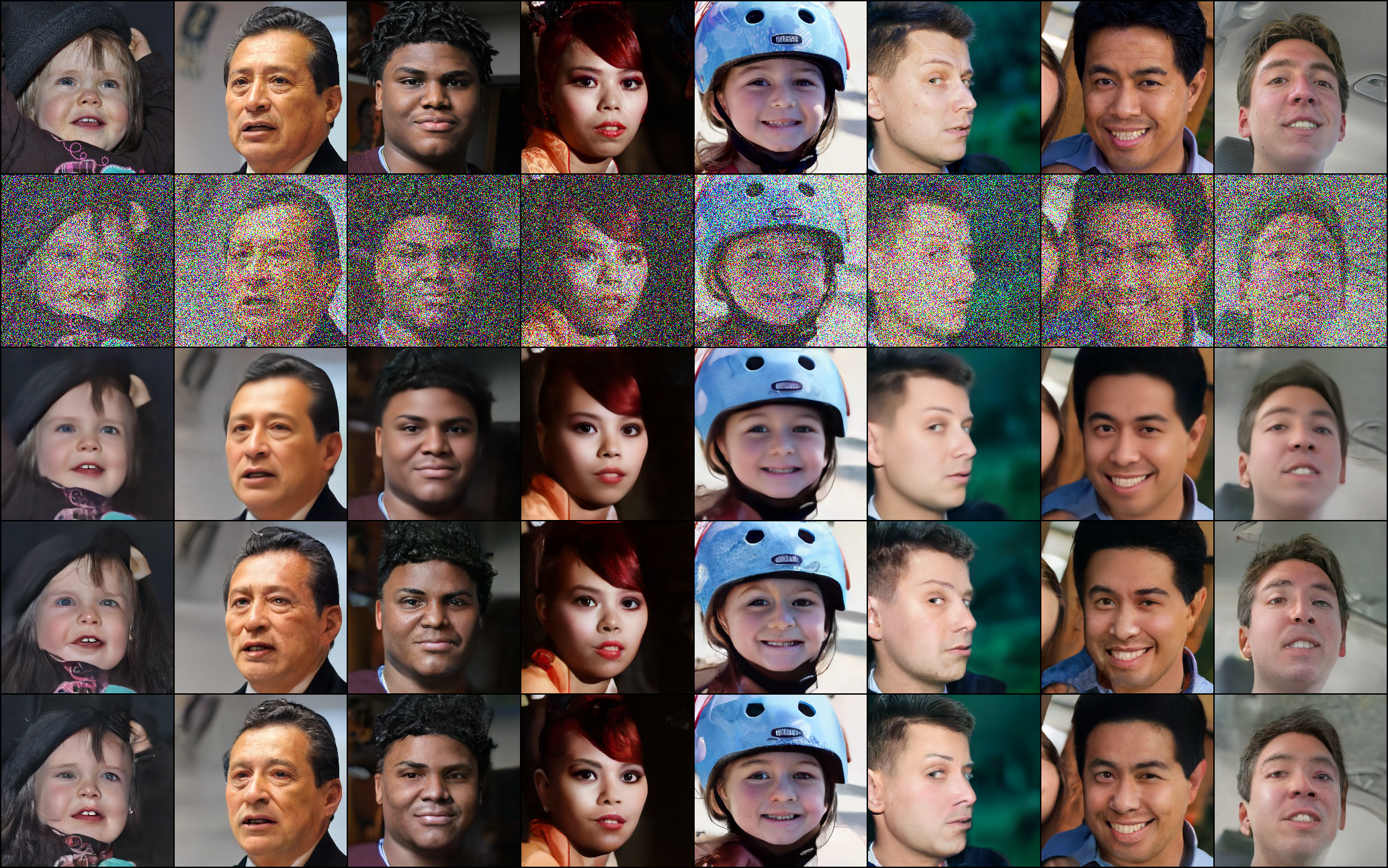}
    \caption{From top to bottom: original FFHQ images, noisy versions with $\sigma_0=0.602$, MMSE denoiser outputs, and two instances of our denoising algorithm's output.}
\end{figure*}
\begin{figure*}
    \centering
    \includegraphics[width=\linewidth]{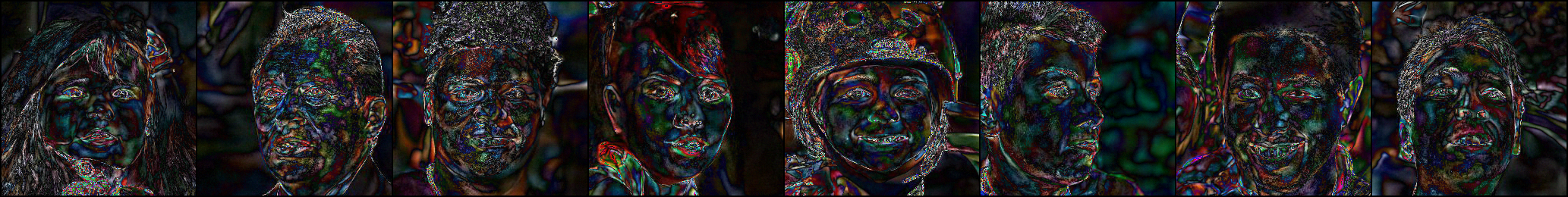}
    \caption{The absolute value of the difference (scaled by $4$) between two instances of our denoising algorithm's output on FFHQ images with $\sigma_0=0.602$.}
\end{figure*}
\begin{figure*}
    \centering
    \includegraphics[width=\linewidth]{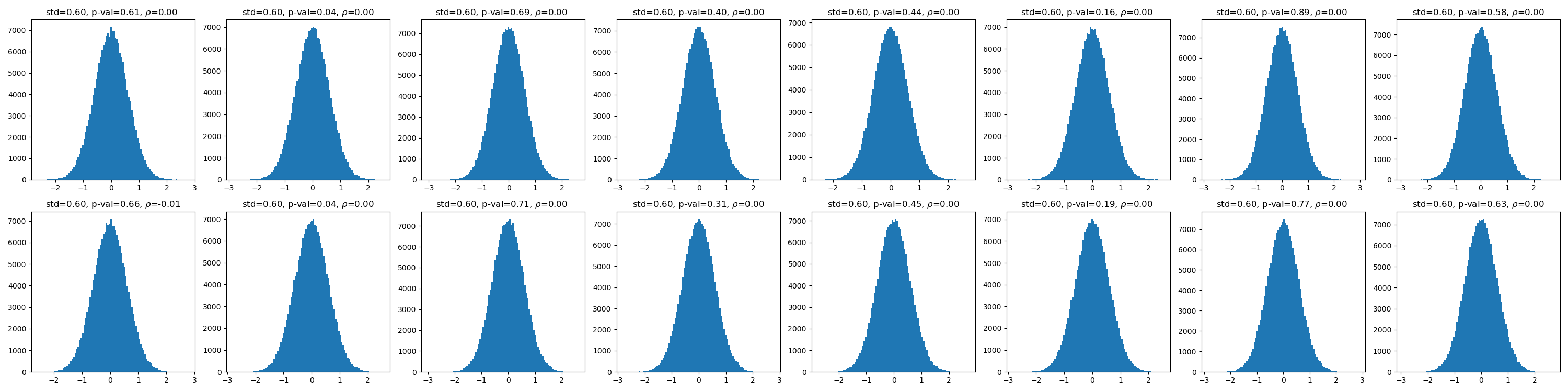}
    \caption{Residual histograms, standard deviations, normality p-values, and the Pearson's correlation coefficients (in the direction with the maximum absolute value) for our denoising algorithm's output on FFHQ images with $\sigma_0=0.602$.}
\end{figure*}

\begin{figure*}
    \centering
    \includegraphics[width=\linewidth]{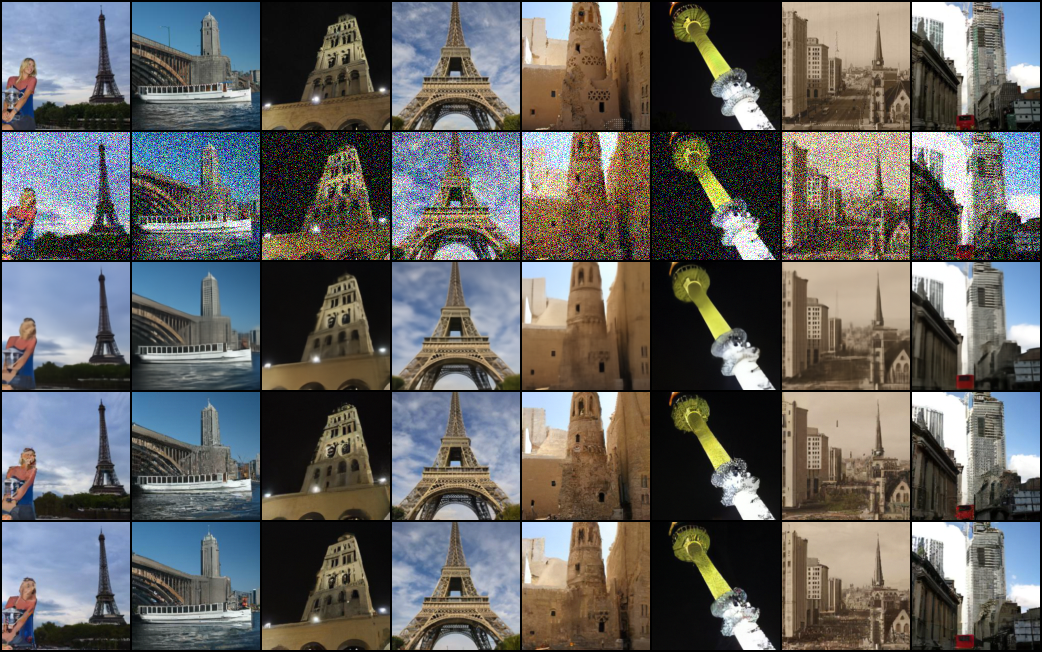}
    \caption{From top to bottom: original LSUN-tower images, noisy versions with $\sigma_0=0.198$, MMSE denoiser outputs, and two instances of our denoising algorithm's output.}
\end{figure*}
\begin{figure*}
    \centering
    \includegraphics[width=\linewidth]{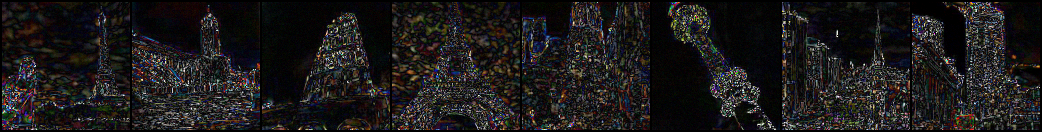}
    \caption{The absolute value of the difference (scaled by $4$) between two instances of our denoising algorithm's output on LSUN-tower images with $\sigma_0=0.198$.}
\end{figure*}
\begin{figure*}
    \centering
    \includegraphics[width=\linewidth]{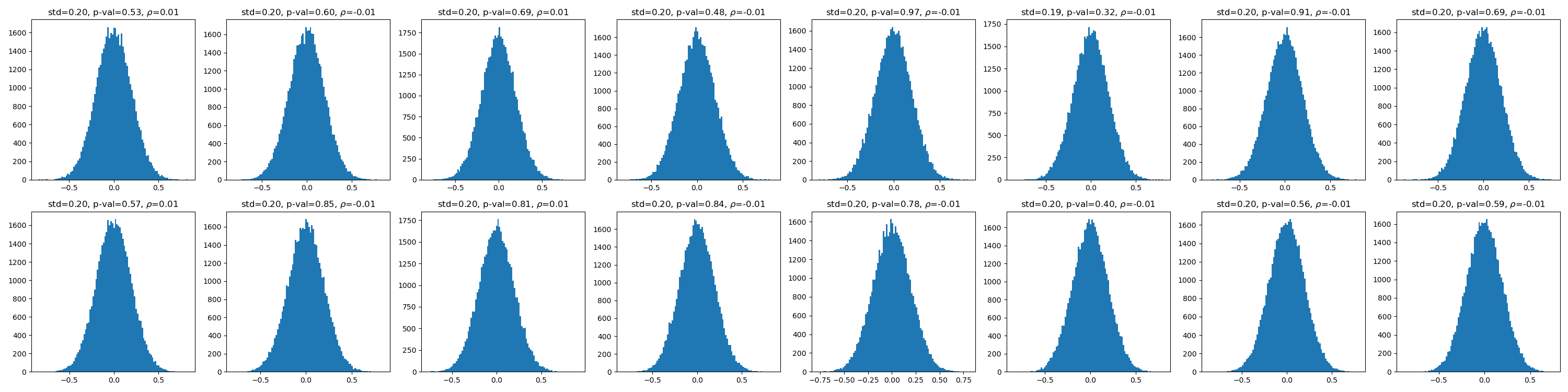}
    \caption{Residual histograms, standard deviations, normality p-values, and the Pearson's correlation coefficients (in the direction with the maximum absolute value) for our denoising algorithm's output on LSUN-tower images with $\sigma_0=0.198$.}
\end{figure*}

\begin{figure*}
    \centering
    \includegraphics[width=\linewidth]{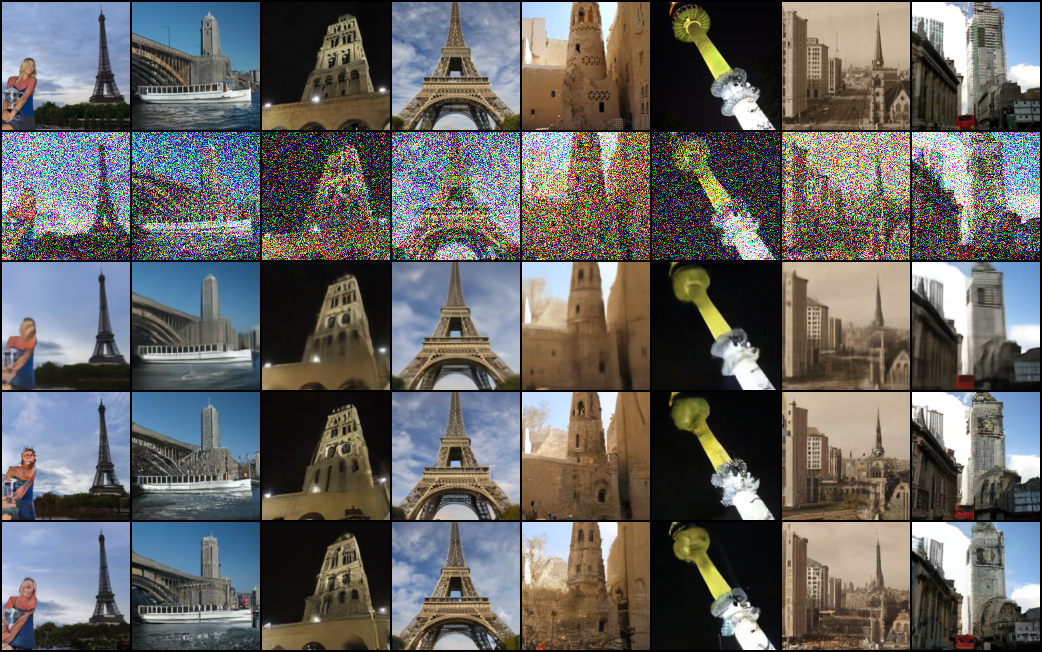}
    \caption{From top to bottom: original LSUN-tower images, noisy versions with $\sigma_0=0.403$, MMSE denoiser outputs, and two instances of our denoising algorithm's output.}
\end{figure*}
\begin{figure*}
    \centering
    \includegraphics[width=\linewidth]{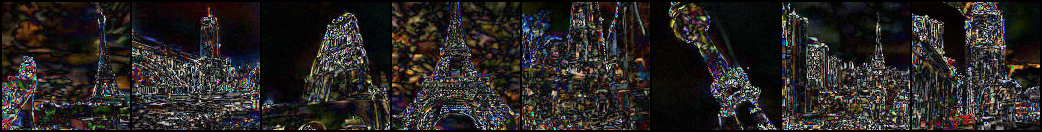}
    \caption{The absolute value of the difference (scaled by $4$) between two instances of our denoising algorithm's output on LSUN-tower images with $\sigma_0=0.403$.}
\end{figure*}
\begin{figure*}
    \centering
    \includegraphics[width=\linewidth]{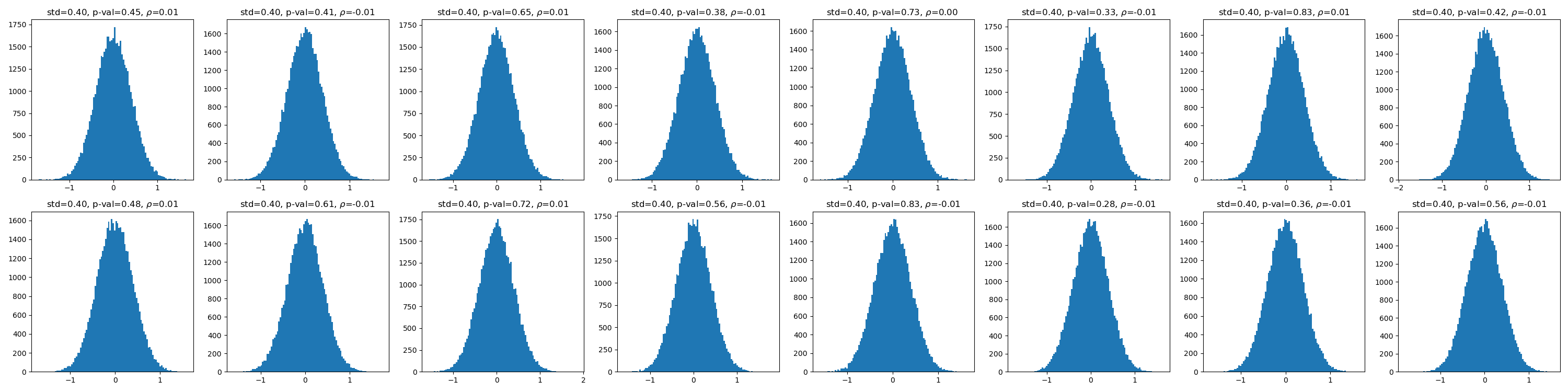}
    \caption{Residual histograms, standard deviations, normality p-values, and the Pearson's correlation coefficients (in the direction with the maximum absolute value) for our denoising algorithm's output on LSUN-tower images with $\sigma_0=0.403$.}
\end{figure*}

\begin{figure*}
    \centering
    \includegraphics[width=\linewidth]{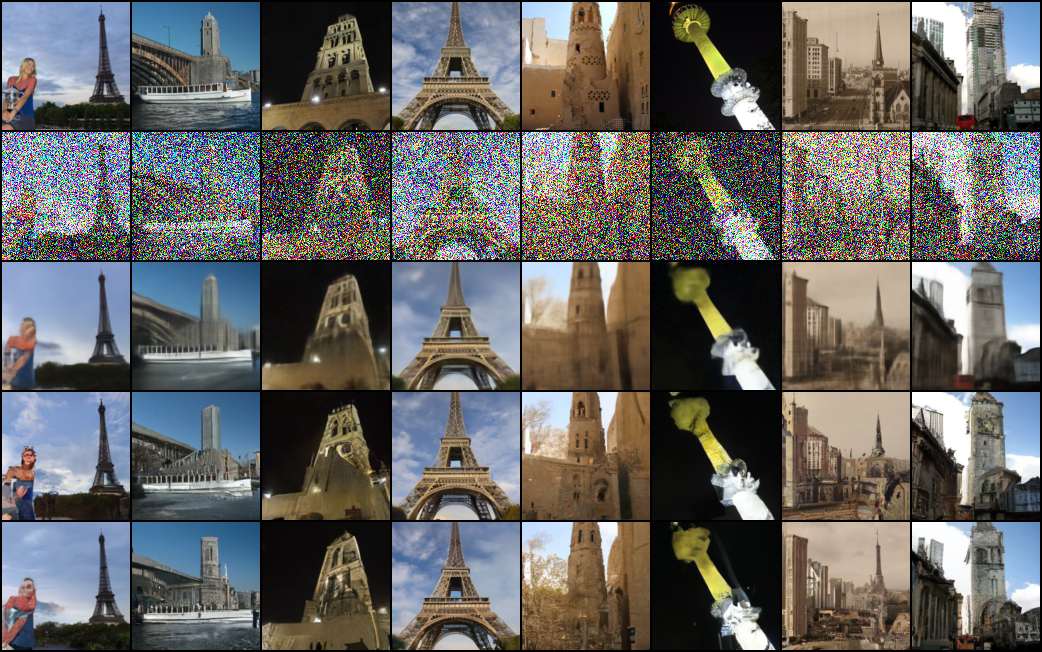}
    \caption{From top to bottom: original LSUN-tower images, noisy versions with $\sigma_0=0.606$, MMSE denoiser outputs, and two instances of our denoising algorithm's output.}
\end{figure*}
\begin{figure*}
    \centering
    \includegraphics[width=\linewidth]{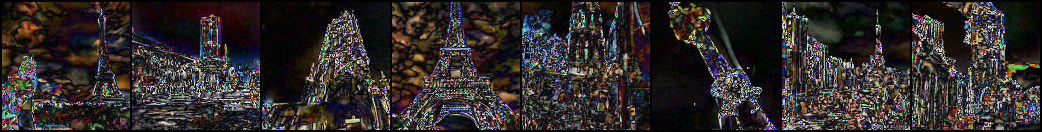}
    \caption{The absolute value of the difference (scaled by $4$) between two instances of our denoising algorithm's output on LSUN-tower images with $\sigma_0=0.606$.}
\end{figure*}
\begin{figure*}
    \centering
    \includegraphics[width=\linewidth]{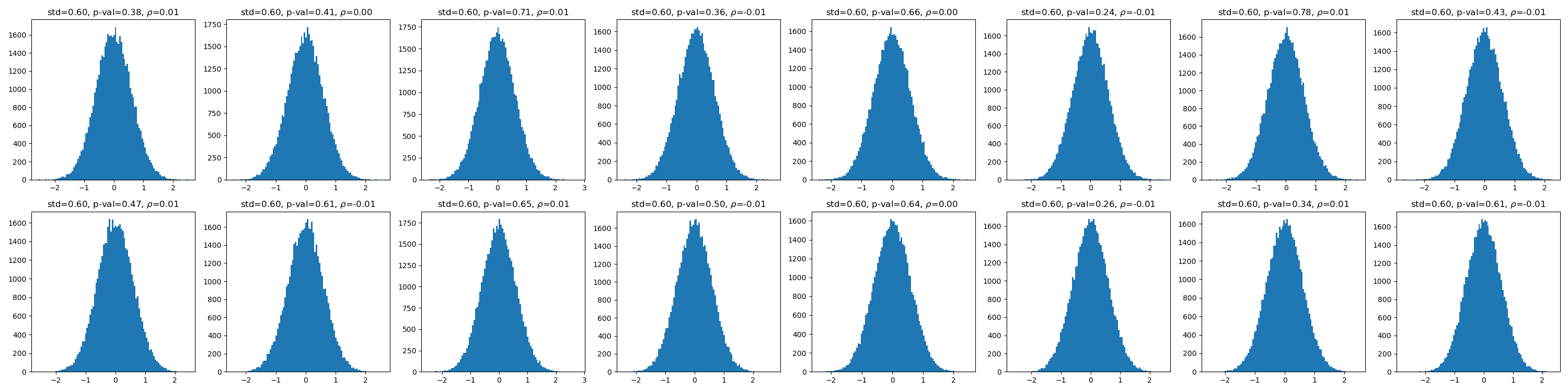}
    \caption{Residual histograms, standard deviations, normality p-values, and the Pearson's correlation coefficients (in the direction with the maximum absolute value) for our denoising algorithm's output on LSUN-tower images with $\sigma_0=0.606$.}
\end{figure*}

\begin{figure*}
    \centering
    \includegraphics[width=\linewidth]{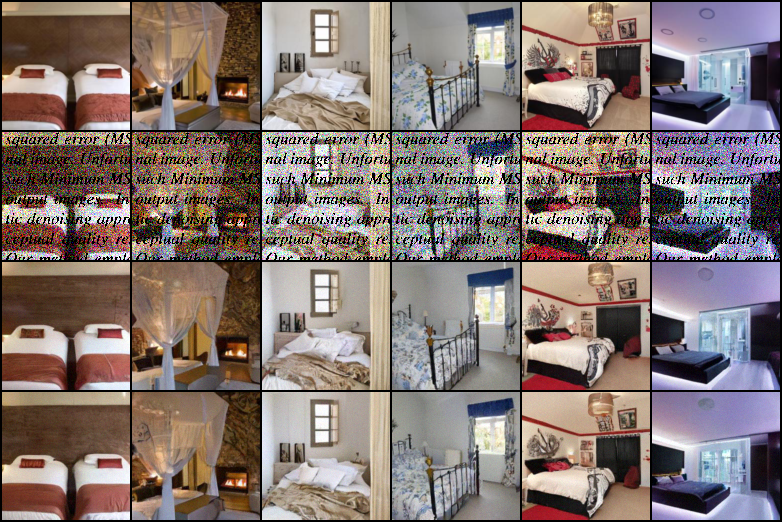}
    \caption{From top to bottom: original LSUN-bedroom images, the observations with text overlay and additive noise ($\sigma_0 = 0.198$), and two instances of our inpainting algorithm's output.}
\end{figure*}

\end{appendices}

\end{document}